\documentclass[twocolumn]{aastex63}
\usepackage{natbib}
\usepackage{appendix}
\usepackage{multirow}
\usepackage{longtable}
\usepackage{afterpage}
\usepackage{color,soul}
\newcommand{\tnt}{\textsc{Dynamite}}
\newcommand{\hd}{\object{HD\,219134}}

\received{}
\revised{}
\accepted{}
\submitjournal{AJ}

\shorttitle{HD\,219134}
\shortauthors{Dietrich et al.}
\graphicspath{{./}{}}
\begin{document}

\defcitealias{Motalebi2015}{M15}
\defcitealias{Vogt2015}{V15}
\defcitealias{Gillon2017a}{G17}

%\title{DYNAMITE Integrated Analysis including Inclination, Eccentricity, and a New Dynamical Stability Criterion, with Predictions on the \hd{} Planetary System}

\title{An Integrative Analysis of the \hd{} Planetary System and the Inner Solar System: Extending DYNAMITE with Enhanced Orbital Dynamical Stability Criteria}

\correspondingauthor{Jeremy Dietrich}
\email{jdietrich1@email.arizona.edu}

\author[0000-0001-6320-7410]{Jeremy Dietrich}
\affiliation{Department of Astronomy, The University of Arizona, Tucson, AZ 85721, USA}

\author[0000-0003-3714-5855]{D\'aniel Apai}
\affiliation{Department of Astronomy, The University of Arizona, Tucson, AZ 85721, USA}
\affiliation{Lunar and Planetary Laboratory, The University of Arizona, Tucson, AZ 85721, USA}

\author[0000-0002-1226-3305]{Renu Malhotra}
\affiliation{Lunar and Planetary Laboratory, The University of Arizona, Tucson, AZ 85721, USA}

%% Note that the \and command from previous versions of AASTeX is now
%% depreciated in this version as it is no longer necessary.  AASTeX 
%% automatically takes care of all commas and "and"s between authors names.

%% AASTeX 6.3 has the new \collaboration and \nocollaboration commands to
%% provide the collaboration status of a group of authors.  These commands 
%% can be used either before or after the list of corresponding authors.  The
%% argument for \collaboration is the collaboration identifier.  Authors are
%% encouraged to surround collaboration identifiers with ()s.  The 
%% \nocollaboration command takes no argument and exists to indicate that
%% the nearby authors are not part of surrounding collaborations.

%% Mark off the abstract in the ``abstract'' environment.  
\begin{abstract}
Planetary architectures remain unexplored for the vast majority of exoplanetary systems, even among the closest ones, with potentially hundreds of planets still ``hidden" from our knowledge.  \tnt{} is a powerful software package that can predict the presence and properties of these yet undiscovered planets.  We have significantly expanded the integrative capabilities of \tnt{}, which now allows for (i) planets of unknown inclinations alongside planets of known inclinations, (ii) population statistics and model distributions for the eccentricity of planetary orbits, and (iii) three different dynamical stability criteria.  We demonstrate the new capabilities with a study of the \hd{} exoplanet system consisting of four confirmed planets and two likely candidates, where five of the likely planets are Neptune-size or below with orbital periods less than 100 days.  By integrating the known data for the \hd{} planetary system with contextual and statistical exoplanet population information, we tested different system architecture hypotheses to determine their likely dynamical stability.  Our results provide support for the planet candidates, and we predict at least two additional planets in this system. We also deploy \tnt{} on analogs of the inner Solar System by excluding Venus or Earth from the input parameters to test \tnt{}'s predictive power.  Our analysis finds the system remains stable while also recovering the excluded planets, demonstrating the increasing capability of \tnt{} to accurately and precisely model the parameters of additional planets in multi-planet systems.
\end{abstract}

%% Keywords should appear after the \end{abstract} command.  
%% See the online documentation for the full list of available subject
%% keywords and the rules for their use.

%% From the front matter, we move on to the body of the paper.
%% Sections are demarcated by \section and \subsection, respectively.
%% Observe the use of the LaTeX \label
%% command after the \subsection to give a symbolic KEY to the
%% subsection for cross-referencing in a \ref command.
%% You can use LaTeX's \ref and \label commands to keep track of
%% cross-references to sections, equations, tables, and figures.
%% That way, if you change the order of any elements, LaTeX will
%% automatically renumber them.
%%
%% We recommend that authors also use the natbib \citep
%% and \citet commands to identify citations.   The citations are
%% tied to the reference list via symbolic KEYs.  The KEY corresponds
%% to the KEY in the \bibitem in the reference list below.

\section{Introduction} \label{sec:intro}

Understanding the stability of planetary systems and searching for a form of order in their dynamics has long been a path of study for astronomers, first for our own Solar System and then in other systems discovered over the past three decades.  While we know much about exoplanets as a population, thanks to the transit survey conducted by the Kepler Space Telescope \citep[][]{Borucki2010} as well as various ground-based radial velocity (RV) surveys, certain questions regarding the dynamical properties of exoplanet systems still remain to this day.  The underlying physics is complex and time-consuming to compute mathematically, and astronomers often have to make a compromise between accuracy and speed when modeling these interactions.  Still, previous studies have been able to ascertain simple empirical rules for orbital dynamical stability that fit the known population statistics fairly well \citep[e.g.,][]{Pu2015, Gilbert2020} and detailed computational models that sacrifice speed for more accuracy via N-body integrations, which work well for the analysis of a single, dynamically complicated system \citep[][]{Volk2020}.

The \tnt{}\footnote{\url{https://github.com/JeremyDietrich/dynamite}} \citep[][]{Dietrich2020} software package provides a unique capability for integrated analysis to combine specific (but often uncertain and incomplete) data of an exoplanet system with population-level statistical information and an empirical rule or other criterion for orbital dynamical stability.  This analysis then predicts the presence and parameters of additional ``hidden" planets in these planetary systems, which provides enhanced statistical understanding of exoplanet populations.  Multiple studies have used \tnt{} to predict the positions, orbits, and properties (size, nature) of additional planets in dozens of multi-planet systems \citep[e.g.,][Basant et al. 2021, submitted]{Dietrich2020,Dietrich2021,HardegreeUllman2021}.  In addition, predictions from \tnt{} have matched known planets excluded from the analysis, as well as new planets found at the time the integrated analysis was performed (Peterson et al. 2021, submitted).  In previous versions of \tnt{}, we adopted a simple dynamical stability criterion from \citep[][]{He2019}, which was guided by N-body simulations of multi-planet dynamical stability to find that most planet pairs are stable only when their mutual Hill radii is $\gtrsim$ 8.  However, this criterion, while useful, does not account for some important details - how would the deviation of an orbit from circular along the plane of the system affect the stability? What about orbital resonances with other planets in the system?  Which methods to assess stability provide the best mix of accuracy and speed?  This paper explores these questions and evaluates three different models of dynamical stability of multi-planet systems.

The \hd{} system, which is the closest star to the Solar System with at least 6 planets or planet candidates and the closest system with transiting planets \citep[][hereafter \citetalias{Motalebi2015,Vogt2015,Gillon2017a}]{Motalebi2015,Vogt2015,Gillon2017a}, provides a complex, high-multiplicity system on which the various stability analyses can be tested with \tnt{}.  With two planets confirmed to transit by the Transiting Exoplanet Survey Satellite \citep[TESS;][]{Ricker2015} and two of the other radial velocity signals not fully confirmed by various studies, we can also qualitatively assess support for the candidate planets, along with searching for additional planets that are still unknown.  The \hd{} system will likely be a prime target for direct imaging missions and next-generation ground-based telescopes in the near future, so having a more complete understanding of the planetary system through the predictions we can generate from our \tnt{} analysis can help guide those observations and instruments.

In addition, the Solar System has provided a great laboratory for analyzing dynamical stability models, including the effects of general relativity on Mercury's orbit.  Thus, as a benchmark, we also deploy \tnt{} to analyze the results from the stability criteria on the inner Solar System (up to and including Jupiter) as an analog for exoplanet systems, including assessing the system with the exclusion of one of the inner planets.  The level of understanding of the Solar System will give strong support on the robustness of our stability measures, as well as continuing to develop the knowledge on how unique the Solar System's architecture truly is \citep[e.g.,][]{Mulders2018}.

This paper is organized as follows. Section~\ref{sec:system} discusses the current knowledge of the \hd{} system, the closest system with 5 or more planets.  Section~\ref{sec:analysis} introduces the new features of the \tnt{} software package.  The results of the new analysis on the \hd{} system are shown in Section~\ref{sec:results}.  Finally, Sections~\ref{sec:TNT_disc}-\ref{sec:SS_disc} discuss the effects of our model assumptions and different parameter tests performed on the \hd{} system, the dynamical stability criteria evaluated, and their different use cases, and a comparison case via modeling the inner Solar System, including Jupiter.

\section{The HD 219134 system} \label{sec:system}

\hd{} (also known as Gliese\,892 or HR\,8832) is a main-sequence early-K dwarf, approximately 80\% the size of the Sun and about four times less luminous \citep[][]{Gray2003,Takeda2007,Boyajian2012}.  It is located at a distance of $6.532\pm0.004$ pc from us \citep[][]{Gaia2018}, making it one of the 30 closest known planet-hosting stars to our Solar System and the closest to host at least 5 planets.  \hd{} is known to be an impressively old star, with derived age estimates of $11.0\pm2.2$ Gyrs \citepalias[][]{Gillon2017a} and $12.46\pm0.5$ Gyrs  \citep[][]{Takeda2007} from various stellar evolution models.  The rotation period of the star, via $v \sin i$ measurements and stellar activity measurements taken over 12 years, is likely either $P_{rot} \approx 22.8$ days or $P_{rot} \approx 42.3$ days, relatively short for such an aged star \citepalias[\citeauthor{Johnson2016}~\citeyear{Johnson2016};][]{Vogt2015, Gillon2017a}.  The stellar parameters for \hd{} can be found in Table~\ref{tab:stellar}.

\begin{table}[t]
    {\centering
    \caption{Stellar parameters for \hd{}}
    \label{tab:stellar}
    \begin{tabular}{|l|c|r|}
        \hline
        \textbf{Parameter Name} & \textbf{Value} & \textbf{Ref.}\\
        \hline
        Spectral Type & K3V & (a)\\
        \hline
        Mass ($M_\odot$) & $0.81\pm0.03$ & (b)\\
        \hline
        Radius ($R_\odot$) & $0.778\pm0.004$ & (c)\\
        \hline
        Luminosity ($L_\odot$) & $0.265\pm0.002$ & (c)\\
        \hline
        Temperature (K) & $4699\pm16$ & (c)\\
        \hline
        Distance (pc) & $6.532\pm0.004$ & (d)\\
        \hline
        \multirow{2}{*}{Rotation period (days)} & $42.3 \pm 0.1$ & (e)\\
        & $22.8\pm0.03$ & (f)\\
        \hline
        \multirow{2}{*}{Age (Gyr)} & $11.0\pm2.2$ & (b)\\
        & $12.46 \pm 0.5$ & (g)\\
        \hline
    \end{tabular}
    }
    \\[10pt]
    \textbf{Notes}: (a) \citet[][]{Gray2003}, (b) \citetalias[][]{Gillon2017a}, (c) \citet[][]{Boyajian2012}, (d) \citet[][]{Gaia2018}, (e) \citetalias[][]{Motalebi2015}, (f) \citet[][]{Johnson2016}, (g) \citet[][]{Takeda2007}.  
\end{table}

\begin{figure*}[ht]
    \centering
    \includegraphics[width=2.12\columnwidth]{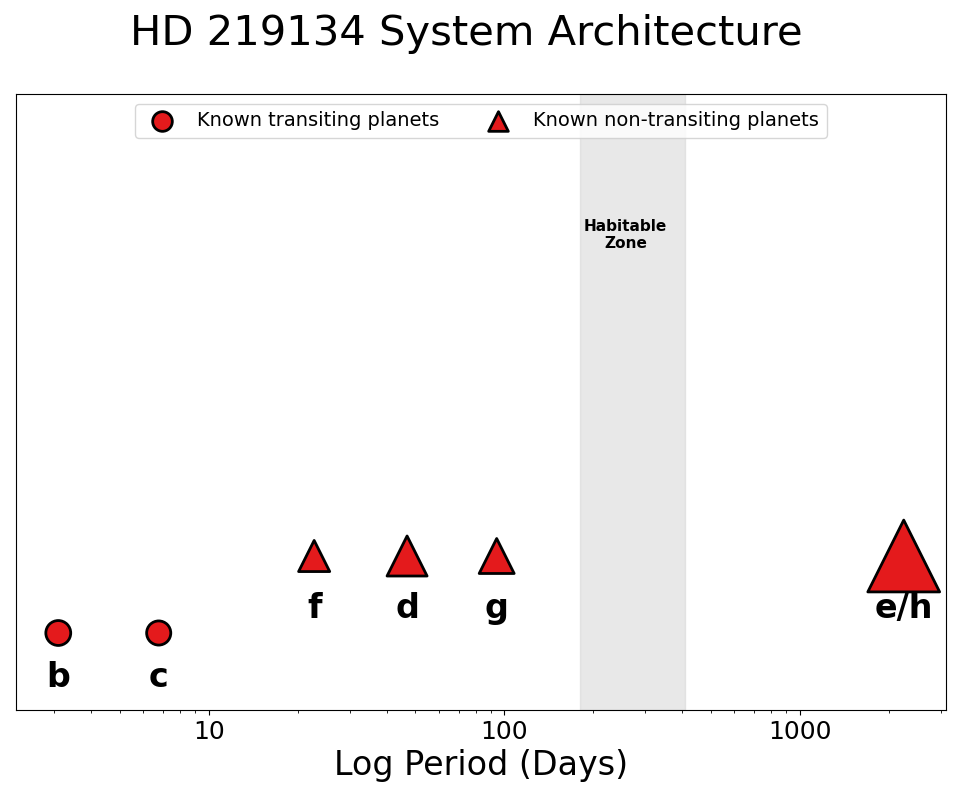}
    \caption{The \hd{} system architecture with known planets and the extent of the habitable zone.  Relative marker sizes match planet sizes.  The label ``e/h" refers to planet e and planet h likely being the same planet, as they were roughly the same mass but with $\sim$20\% difference in orbital periods.}
    \label{fig:arch}
\end{figure*}

The known system architecture of \hd{} is illustrated in Figure~\ref{fig:arch}.  The first study to detect the presence of planets around \hd{} was done by \citetalias[][]{Motalebi2015}, who found four planets in the system using radial velocity (RV) observations.  Three were sub-Neptunian planets located close to the host star with orbital periods of 3.094 (b), 6.767 (c), and 46.66 (d) days, whereas the fourth (e) was a Saturnian or Jovian planet much further out with an orbital period of 1842 days.  Shortly thereafter, another RV study by \citetalias[][]{Vogt2015} found evidence for six planets in the the system.  Notably, they confirmed the presence of planets b-d, discovered evidence for two more sub-Neptunian planets in the inner system at orbital periods of 22.81 (f) and 94.2 (g) days, but did not find planet e; however, they found a similarly massive planet (h) in the outer system, but at a noticeably longer period of 2247 days.

Another study by \citet[][]{Johnson2016}, who gathered stellar activity data over a 12 year baseline, found evidence for a rotation period of 22.8 days, while providing no confirmation on the presence of planet f at a very similar orbital period to the rotational/stellar activity period.  However, they did confirm planets d and h, stating that planet e was likely a mis-identification of planet h.  In addition to their RV observations, \citetalias[][]{Motalebi2015} also found that planet b transited \hd{} using photometry from the \textit{Spitzer Space Telescope}.  A follow-up transit photometry analysis also using the \textit{Spitzer Space Telescope} by \citetalias[][]{Gillon2017a} discovered that planet c also transited the host star.  This study determined that by geometric probability alone planet c only had a 5.4\% chance of transiting the host star, but after confirmation of the transits of planet b the transiting probability of planet c increased to 21\% before it too was found to transit.  The observations of planet c transiting, therefore, increase the transit probability of planets f and d to 13\% (from 2.5\%) and 8.1\% (from 1.5\%), respectively, as the additional transiting planet means the system inclination is likely closer to edge-on.

TESS observed the \hd{} system in Sectors 17 and 24 as TOI 1469, and confirmed the transit events for planets b and c but did not detect transits for any other planet\footnote{\url{https://exofop.ipac.caltech.edu/tess/target.php?id=283722336}}.  Thus, it is likely that any planets in the inner system with an orbital period within 30 days would either have an inclination far enough away from edge-on such that they would not transit \hd{} or would be smaller than any known planet in the system.  However, due to TESS's observational strategy, the parameter constraints are not tight for planets with orbital periods longer than TESS's single-sector observing window of $\sim$27 days.

Thus, the presence of planets b-d and e/h have been confirmed in multiple studies, with planet e from the original \citetalias[][]{Motalebi2015} study found to likely be another manifestation of planet h from \citetalias[][]{Vogt2015}.  The validity of planets f and g (both found by \citetalias[][]{Vogt2015}) have not yet been established.  A recent multi-decade analysis of RV data for the \hd{} system by \citet[][]{Rosenthal2021} confirmed a signal at $22.795^{+0.005}_{-0.006}$ days that would correspond to planet f, with a slightly smaller $m \sin i$ value.  However, they did not report a signal near the originally given period of planet g at $\sim$94 days, but analyzed signals near 2$\times$ and 4$\times$ that period and determined them to be false positives due to annual/instrumental systematics (see Section \ref{subsec:add}).

In our analysis of the \hd{} system, we do not take the outer gas giant planet e/h \citepalias[][]{Motalebi2015, Vogt2015} into account, as it is beyond the limits from our current population statistics.  Moreover, its existence and properties are inconsistently reported in different studies, as it is not mentioned in any other studies besides these two discovery papers, which have a relatively small but significant inconsistency in the reported values.  However, future studies should consider its effects on the inner system and its dynamical stability.

We gathered our information on \hd{} from the NASA Exoplanet Archive\footnote{\url{https://exoplanetarchive.ipac.caltech.edu/}} and the associated references above.  Specifically, the specific data we input to \tnt{} are only the planet parameters (orbital periods, radii/m sin i values, inclination, and eccentricity as well as their uncertainties) as they are given from the previous studies; we do not re-analyze any of their datasets as there is no reason to do so.  A detailed summary of the exoplanet population statistical information is detailed in \citet[][]{Dietrich2020, Dietrich2021}, and we only note here that the key trends follow those reported in \citep[][]{Fabrycky2014, Mulders2018, He2019, He2020}.  In order to provide a complete analysis despite the presence of planets f and g not being well-established in the previous datasets, we also explore the different configurations of the system if planets f and/or g are considered ``unconfirmed" and are not included in the system architecture.

\section{Analysis} \label{sec:analysis}

\subsection{The \tnt{} Software Package} \label{subsec:dynamite}

Here we utilize and expand upon \tnt{} for our integrated analysis \citep[][]{Dietrich2020}.  We use \tnt{} to perform a unique integrated analysis of the specific yet incomplete data currently gathered from observations of a system, combined with an orbital dynamical stability criterion for the system and population statistical models for orbital periods, planet radii, inclination, and planet mass-radius relationships. From this analysis, we give predictions on the most likely values for the parameters of one additional planet in the system.  Fundamentally, \tnt{} provides a quantitative answer to the following question: given what we know about the exoplanet population and about this specific planetary system, if there is one yet-undiscovered planet in this system, what are its most likely parameters?

The likelihood distributions for these predictions are derived from the Kepler population of planets \citep[see e.g.,][]{Mulders2018, He2019, He2020}, and the constraints on our predictions are reliant on the prior information given.  These priors are gathered from thousands of systems, and therefore are most accurate for systems where the planetary architecture is most similar to those in typical exoplanet systems from the population.  If a specific system is exceptional, our predictions may not be accurate and could be contradicted by observations, alerting us to the outlier status of that system.  Also, additional priors such as higher precision for data or additional demographic information (e.g., occurrence rate by spectral type, stellar rotational period, etc.) will continue to refine the assumptions in our model (or the specific hypotheses tested) and will adjust our predictions accordingly.

\tnt{} requires as input the orbital periods and radii or minimum masses and their uncertainties (depending on availability from different types of observations) of the planets, as well as the radius, mass, and luminosity of the host star.  Additional parameters -- where unknown values are allowed -- include the inclination and eccentricity of the planets.  \tnt{} then calculates the statistical likelihood distributions given the population models (of which there are multiple modules for the different parameters that can be switched and tested separately) and runs a Monte Carlo sampling chain on them.  Every set of parameters is tested for dynamical stability and discarded if the system would be considered dynamically unstable (see \S\ref{subsec:dynstab} for our new implementation on this consideration).

For ranges in the orbital period distribution that are stable, the likelihood is determined based on the orbital period module chosen.  Currently, the two available modules are ``equal period ratios", where planets are most likely to be found when their pairwise period ratios are similar \citep[][]{Mulders2018}, and ``clustered periods", where planets are most likely to be found in clusters with potential gaps in the system \citep[][]{He2019}.  We label peaks in the orbital period distribution as significant when $\gtrsim$ 90\% of injections interior to the outermost planet are found in a specific location in the space.  We also note positions where multiple planets could exist if we were testing for the presence of more than one additional planet.  The output data of \tnt{} are testable predictions in the form of orbital period, transit depth/probability, RV semi-amplitude, and direct imaging separation and contrast for the planet parameters with the highest likelihood.

The method was originally tested on known Kepler systems with planets removed; such planets were recovered with a good degree of accuracy on the parameters.  \tnt{} was developed on the known TESS multi-planet systems and refined further for analyzing the $\tau$ Ceti planetary system, where we predict at least one possibly rocky habitable-zone planet is likely to exist but has not yet been discovered \citep[][]{Dietrich2021}.  Additional studies have utilized \tnt{} analyses for the K2-138 system with at least 6 planets \citep[][]{HardegreeUllman2021}, as well as both Mars in the Solar System and the e Eridani (also known as HD 20794; not to be confused with $\epsilon$ Eridani) system of 3 known planets and 3 planet candidates (Basant et al. 2021, submitted).  Here we will briefly discuss the objectives of \tnt{}, in the context of the significant extension of functionalities we added for the present study.

\begin{figure*}[ht]
    \centering
    \includegraphics[width=2.1\columnwidth]{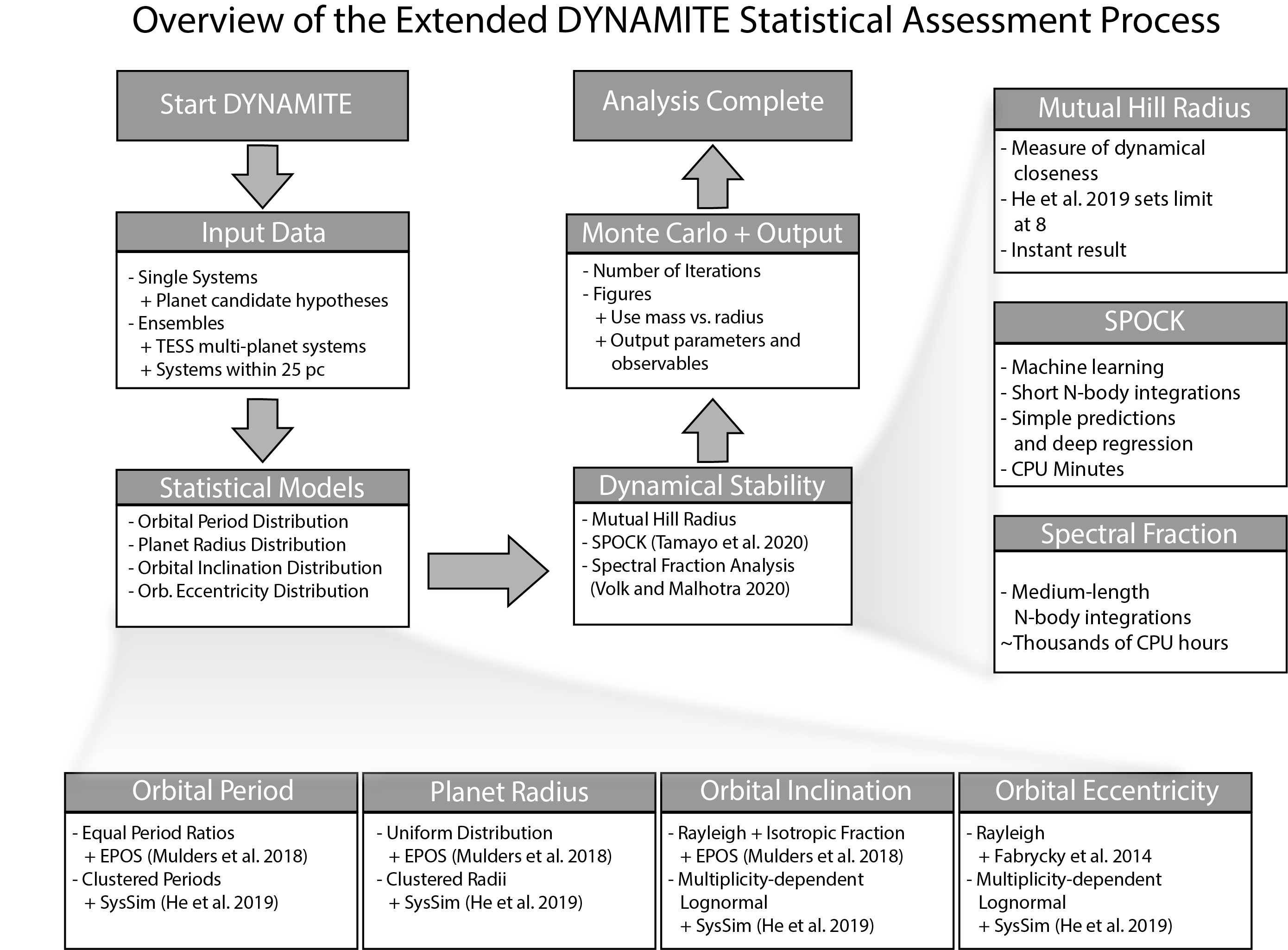}
    \caption{The flowchart of \tnt{}, with the model choices that are set up in the program's configuration file.}
    \label{fig:flow}
\end{figure*}

Our goal is to utilize the combined statistical knowledge gathered across all exoplanet systems to determine likelihoods for the presence and properties of planets in specific systems.  Our analysis is, therefore, reliant on the accuracy and precision of the data we can gather for a system, as well as our ability to model system parameters and orbital stability.  We run our analysis in multiple modes with these different modules of data models to assess the accuracy of these models for future analyses and to provide testable hypotheses for models with uncertain levels of knowledge.  \tnt{} has different models for the orbital periods, planet radii, mass-radius relationship, orbital inclination, and orbital eccentricity, and each combination of models provides a different set of predicted planet injections with different likelihoods.  The choices made by our analysis are shown in a flowchart in Figure~\ref{fig:flow}.  Our motivation to upgrade \tnt{} and expand its capabilities was to determine the fine structure in dynamical orbital stability, as our previous simple stability model was sufficient for first-approximation results but is likely not a precise model.  Here we discuss the updated models and parameter decisions added to \tnt{}.

\subsection{Planet existence hypotheses} \label{subsec:hypo}

\begin{figure*}[ht]
    \centering
    \includegraphics[width=2.12\columnwidth]{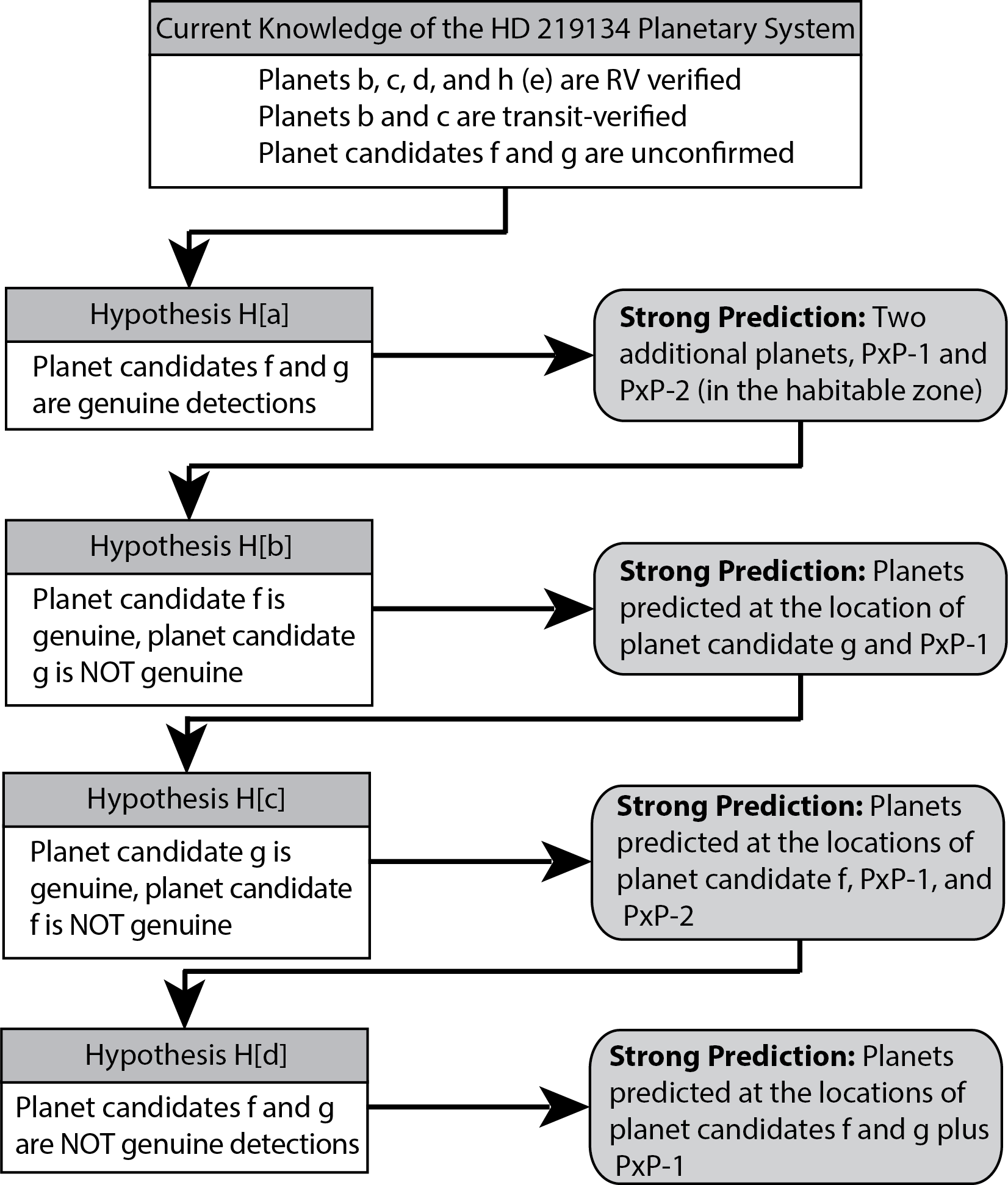}
    \caption{The decision tree of data for \tnt{}, showing the different hypotheses that can be chosen for the \hd{} system and can be updated for any given exoplanet system.}
    \label{fig:tree}
\end{figure*}

We test multiple different hypotheses of the system architecture to determine how incorporating the presence of not-yet-confirmed or uncertain planets and planet candidates changes our understanding of the system architecture and planet properties.  Each of these hypotheses is a ``branch" on the \tnt{} data decision tree (see Figure~\ref{fig:tree}) and we explore how these decisions change the results.

In the \hd{} system, there are two planets with some degree of uncertainty about their existence.  Planet f has an orbital period of 22.7 days, which is very close to a measured rotational period for its host star of 22.8 days \citep[][]{Johnson2016}.  However, it was included as a known planet for the purpose of updating transit probabilities from the Spitzer follow-up transit study by \citetalias[][]{Gillon2017a}, and a signal with the same period was found over long-term RV data by \citet[][]{Rosenthal2021}.  Planet g, reported by \citetalias[][]{Vogt2015}, has no other measurements from any other study; \citetalias{Motalebi2015} and \citetalias[][]{Gillon2017a} do not reference it anywhere in their analysis, while \citet[][]{Johnson2016} briefly mention it but do not provide any additional analysis to confirm or reject its existence.  The parameter values from the analysis by \citetalias[][]{Vogt2015} are much less precise than the other inner planets, but there is no reason to specifically doubt its presence in the system.

For our analysis, we have decided to explore four different system hypotheses in which we include and exclude planet f and planet g from the system architecture, and exclude the outer gas giant planet as stated above.  The four hypotheses are defined as below:
\begin{itemize}
    \item \textbf{Hypothesis H[a]}: All five planets b, c, d, f, and g exist at their current parameter values.
    \item \textbf{Hypothesis H[b]}: The four planets b, c, d, and f exist, while planet g does not.
    \item \textbf{Hypothesis H[c]}: The four planets b-d and g exist, while planet f does not.
    \item \textbf{Hypothesis H[d]}: The three planets b-d exist, while planets f and g do not.
\end{itemize}
We treat the excluded planets as ``unconfirmed" in our analysis and are not known by the injection algorithm.  If the Monte Carlo analysis returns a high relative likelihood at the same orbital periods, we consider this as providing support to the existence of these planets.

\subsection{Inclination and eccentricity distributions}

We expanded the parameter space available for analysis and refined our modeling by incorporating planets with unknown inclinations and testing two different models for the inclination distribution.  For systems where planetary inclinations are unconstrained, our analysis now assumes an isotropic distribution for the system inclination.  For systems where only some planets have inclination constraints, we fit the known values to the requisite distribution of the system inclination for each model described below.

After setting a system inclination (which varies in the isotropic case for each iteration), we distribute the mutual inclinations in two ways.  The first model comes from the Exoplanet Populations Observation Simulator \citep[EPOS;][]{Mulders2018} and assumes a fraction of systems have isotropically-distributed inclinations, while the rest have a narrow ($\sigma_i \sim 2$) Rayleigh distribution of mutual inclinations.  The second model comes from the Exoplanets Systems Simulator \citep[SysSim;][]{He2019,He2020} and assumes a Lognormal distribution with parameters dependent on the multiplicity of the system, where more planets in the system tends to lead to a more coplanar system.

We also included eccentricity distributions, both as another parameter for which we could assess probabilities and as a necessity for the N-body integrations for the dynamical stability criterion (via orbit crossing determinations and AMD measurements).  A large fraction of the eccentricities in the Kepler multi-planet system population are $\lesssim 0.1$ \citep[e.g.,][]{Xie2016, Mills2019, VanEylen2019}.  Our first eccentricity distribution is a Rayleigh distribution with parameter $\sigma_e = 0.02$ \citep[see e.g.,][]{Fabrycky2014}, which is not dependent on any other parameter in the system.  We also used a multiplicity-dependent Lognormal distribution from SysSim, and as with the inclination model, more planets in the system implies more circular orbits for the planets.  For planets without known eccentricities, we sample them from the corresponding distribution for the multiplicity in the system, given that we have injected one planet into the system.

\subsection{Dynamical stability} \label{subsec:dynstab}

Our goal is to assess the precision of our dynamical stability model and compare the trade-offs required between complexity and speed.  The dynamical stability criterion determines the likelihood that the parameters of the predicted planets injected into the system by \tnt{} do not affect the long-term orbital stability of the other planets in the system and of the injected planet itself.  This is an important limiting factor on the injection likelihoods, but it can be a complex and time-consuming process to model.  In particular, orbit crossings (i.e., the inner planet of a planet pair having an apastron larger than the periastron of the outer planet in the pair) are not always an indicator of dynamical instability (e.g., in the Solar System the orbits of Neptune and Pluto cross).  Here, we make the simplifying assumption that orbital crossings are unstable for planets of similar masses on timescales longer than 1 Gyr, since the reported age of the \hd{} system is $>10$ Gyr.  We discuss further the effect of this assumption in Section~\ref{sec:stable_disc}. 

The dynamical stability criterion implemented in our previous \tnt{} analyses was derived from a study by \citet[][]{He2019}, who determined that a simple model where planet pairs within 8 mutual Hill radii were unstable and planet pairs outside that value were stable was a good empirical match to current data.  However, this criterion did not catch finer details relating to orbital resonances and secular chaos in systems.  Therefore, in this study we compare the results from the simple mutual Hill radii cutoff model previously used with two other more detailed analyses involving N-body integrations from \citet[][]{Tamayo2020} and \citet[][]{Volk2020}.

First, we test the Stability of Planetary Orbital Configurations Klassifier \citep[SPOCK;][]{Tamayo2020}.  SPOCK utilizes a machine learning algorithm that trains on 1e5 data sets of planet parameters (both stable and unstable) from N-body integrations and predicts stability using physically motivated summary statistics measured in short N-body integrations ($\sim 10^4$ orbits of innermost planet).  This provides a multiple-order-of-magnitude speed-up over full N-body integrations, and thus can be performed within an hour on a single laptop or desktop machine.  SPOCK provides both a simple predictor based on a single set of summary features and a deep regression analysis that uses thousands of realizations to better characterize the dynamics of the planetary system.  These predictors look at stability of a system as a probabilistic measure instead of the definitive result achieved at the end of the full N-body integration.  The level for which a system becomes unstable is where the predictor's false positive rate hits 10\%, which occurs at a stability threshold of 0.34.  \citet[][]{Tamayo2020} report that SPOCK provides $\gtrsim$90\% accuracy in stability analyses for both predictors, with full N-body simulations being essentially ground truth, and is more accurate in comparison to other stability prediction methods, such as the mutual Hill radius cutoff and the Mean Exponential Growth factor of Nearby Orbits chaos indicator \citep[MEGNO;][]{Cincotta2003}.

Secondly, we test the spectral fraction method from \citet[][]{Volk2020}.  This analysis performs a medium-length ($\sim 5\times10^6$ orbits of innermost planet) N-body integration of the system to determine the likelihood of stability over long N-body integrations ($\sim 5\times10^9$ orbits of innermost planet).  The spectral fraction is determined by taking the power spectrum of the angular momentum deficit \citep[AMD, a measure of the difference in the orbits from circular and coplanar; see e.g.,][]{Laskar2017}.  If a significant fraction of frequencies ($\gtrsim 1\%$) exist at a relatively high percentage of the main peak in the AMD power spectrum ($\gtrsim 5\%$ of the peak), then the system is likely unstable over long ($\sim 5\times10^{9}$ orbits of innermost planet) integrations; otherwise, the system is considered stable.

\section{Results} \label{sec:results}

We report the properties of the planets in the \hd{} system and our predictions for each of our architecture hypotheses, dynamical stability analyses, and statistical model choices in Tables~\ref{tab:planets}-\ref{tab:planets2}.  The figures showing the relative likelihoods for each hypothesis in two dynamical stability tests are shown in Appendix~\ref{app:figures}, with the mutual Hill radius criterion injections in blue (Figure~\ref{fig:simplefg}) and the N-body integration injections with the spectral fraction analysis in green (Figure~\ref{fig:dsc_test1}).  The planet radii are unlikely to be measured via transits for planets external to planet c due to the inclination of the known planets and the increasingly narrow range of transiting inclinations as the semi-major axis increases.  The statistical model choices for the planet radius, inclination, and eccentricity have little effect, whereas the results from each orbital period model (equal period ratios from EPOS \citep[][]{Mulders2018} and clustered periods from SysSim \citep[][]{He2019}) and dynamical stability criterion chosen are similar but do show some significant differences.  We chose to utilize the equal period ratio orbital period model for our analysis.

\begin{table*}[ht]
    \centering
    \caption{Planet and Planet Candidate Periods, Radii, and Masses}
    \label{tab:planets}
    \begin{tabular}{|l|c|c|c|c|r|}
        \hline
        \textbf{Name} & \textbf{Period (days)} & \textbf{Radius ($R_\oplus$)} & \textbf{Mass ($M_\oplus$)} & \textbf{Note} & \textbf{Origin/Reference}\\
        \hline
        \hd{}~b & $3.093 \pm 0.00001$ & $1.602 \pm 0.055$ & $4.74 \pm 0.19$ & Planet & \citetalias[][]{Motalebi2015, Gillon2017a}\\
        \hline
        \hd{}~c & $6.765 \pm 0.0003$ & $1.511 \pm 0.047$ & $4.36 \pm 0.22$ & Planet & \citetalias[][]{Motalebi2015, Gillon2017a}\\
        \hline
        \multirow{8}{*}{H[a] PxP--1} & $12.4\:[10.9,\,14.7]$ & $1.69\: [1.25,\,2.27]$ & $3.99\:[1.99,\,6.36]$ & \multirow{8}{*}{Predicted} & Ratios, ``Volatile", $\Delta = 8$ \\
        & $12.4\:[10.9,\,14.5]$ & $1.79\:[1.33,\,2.41]$ & $6.71\:[2.41,\,18.7]$ & & Ratios, ``Rocky", $\Delta = 8$ \\
        \cline{2-4}
        & $13.3\:[10.8,\,15.4]$ & $1.68\:[1.25,\,2.26]$ & $3.95\:[1.94,\,6.31]$ & & Clustered, ``Volatile", $\Delta = 8$ \\
        & $13.1\:[10.7,\,15.3]$ & $1.77\:[1.32,\,2.39]$ & $6.45\:[2.34,\,18.2]$ & & Clustered, ``Rocky", $\Delta = 8$ \\
        \cline{2-4}
        & $12.4\:[10.7,\,13.7]$ & $1.69\:[1.26,\,2.28]$ & $3.99\:[2.00,\,6.40]$ & & Ratios, ``Volatile", N-body \\
        & $12.4\:[10.7,\,13.5]$ & $1.79\:[1.34,\,2.42]$ & $6.71\:[2.42,\,18.7]$ & & Ratios, ``Rocky", N-body \\
        \cline{2-4}
        & $13.3\:[10.6,\,15.2]$ & $1.68\:[1.26,\,2.27]$ & $3.95\:[1.95,\,6.35]$ & & Clustered, ``Volatile", N-body \\
        & $13.1\:[10.5,\,15.1]$ & $1.77\:[1.33,\,2.40]$ & $6.45\:[2.35,\,18.2]$ & & Clustered, ``Rocky", N-body \\
        \hline
        \multirow{8}{*}{H[b] PxP--1} & \multirow{2}{*}{$12.4\:[10.7,\,14.4]$} & $1.65\:[1.22,\,2.22]$ & $3.84\:[1.84,\,6.14]$ & \multirow{8}{*}{Predicted} & Ratios, ``Volatile", $\Delta = 8$ \\
        & & $1.72\:[1.28,\,2.31]$ & $5.85\:[2.11,\,16.2]$ & & Ratios, ``Rocky", $\Delta = 8$ \\
        \cline{2-4}
        & $13.3\:[10.7,\,15.4]$ & $1.65\:[1.23,\,2.23]$ & $3.84\:[1.79,\,6.09]$ & & Clustered, ``Volatile", $\Delta = 8$ \\
        & $13.2\:[10.7,\,15.4]$ & $1.69\:[1.27,\,2.28]$ & $5.50\:[2.05,\,15.5]$ & & Clustered, ``Rocky", $\Delta = 8$ \\
        \cline{2-4}
        & \multirow{2}{*}{$12.4 [10.7, 13.9]$} & $1.65\:[1.23,\,2.23]$ & $3.84\:[1.84,\,6.18]$ & & Ratios, ``Volatile", N-body \\
        & & $1.72\:[1.29,\,2.32]$ & $5.85\:[2.11,\,16.3]$ & & Ratios, ``Rocky", N-body \\
        \cline{2-4}
        & $13.3\:[10.7,\,15.3]$ & $1.65\:[1.24,\,2.24]$ & $3.84\:[1.79,\,6.13]$ & & Clustered, ``Volatile", N-body \\
        & $13.3\:[10.7,\,15.2]$ & $1.69\:[1.28,\,2.29]$ & $5.50\:[2.05,\,15.6]$ & & Clustered, ``Rocky", N-body \\
        \hline
        \multirow{8}{*}{H[c] PxP--1} & $18.1\:[14.0,\,25.7]$ & $2.13\:[1.30,\,3.82]$ & $5.75\:[2.23,\,14.4]$ & \multirow{8}{*}{Predicted} & Ratios, ``Volatile", $\Delta = 8$ \\
        & $18.1\:[13.9,\,25.7]$ & $1.77\:[1.32,\,2.39]$ & $6.45\:[2.34,\,19.2]$ & & Ratios, ``Rocky", $\Delta = 8$ \\
        \cline{2-4}
        & \multirow{2}{*}{$4.58\:[4.38,\,4.88]$} & $1.91\:[1.25,\,3.62]$ & $4.84\:[1.95,\,13.3]$ & & Clustered, ``Volatile", $\Delta = 8$ \\
        & & $1.70\:[1.28,\,2.30]$ & $5.61\:[2.10,\,15.9]$ & & Clustered, ``Rocky", $\Delta = 8$ \\
        \cline{2-4}
        & $18.1\:[14.2,\,25.2]$ & $2.14\:[1.30,\,3.84]$ & $5.79\:[2.23,\,14.6]$ & & Ratios, ``Volatile", N-body \\
        & $18.1\:[14.1,\,25.2]$ & $1.78\:[1.32,\,2.41]$ & $6.49\:[2.34,\,19.4]$ & & Ratios, ``Rocky", N-body \\
        \cline{2-4}
        & \multirow{2}{*}{$4.58\:[4.38,\,4.88]$} & $1.92\:[1.25,\,3.63]$ & $4.88\:[1.95,\,13.5]$ & & Clustered, ``Volatile", N-body \\
        & & $1.71\:[1.28,\,2.32]$ & $5.65\:[2.10,\,16.0]$ & & Clustered, ``Rocky", N-body \\
        \hline
        \multirow{8}{*}{H[d] PxP--1} & $18.1\:[14.4,\,25.9]$ & $2.05\:[1.29,\,4.04]$ & $5.41\:[2.17,\,15.8]$ & \multirow{8}{*}{Predicted} & Ratios, ``Volatile", $\Delta = 8$ \\
        & $18.1\:[14.5,\,25.7]$ & $1.65\:[1.22,\,2.22]$ & $5.06\:[1.83,\,14.1]$ & & Ratios, ``Rocky", $\Delta = 8$ \\
        \cline{2-4}
        & \multirow{2}{*}{$4.58\:[4.38,\,4.88]$} & $1.82\:[1.23,\,3.66]$ & $4.48\:[1.84,\,13.5]$ & & Clustered, ``Volatile", $\Delta = 8$ \\
        & & $1.59\:[1.20,\,2.09]$ & $4.46\:[1.69,\,11.5]$ & & Clustered, ``Rocky", $\Delta = 8$ \\
        \cline{2-4}
        & $18.1\:[14.2,\,25.4]$ & $2.06\:[1.29,\,4.04]$ & $5.45\:[2.17,\,15.8]$ & & Ratios, ``Volatile", N-body \\
        & $18.1\:[14.3,\,25.2]$ & $1.66\:[1.22,\,2.22]$ & $5.10\:[1.83,\,14.1]$ & & Ratios, ``Rocky", N-body \\
        \cline{2-4}
        & \multirow{2}{*}{$4.58\:[4.38,\,4.88]$} & $1.83\:[1.23,\,3.66]$ & $4.52\:[1.84,\,13.5]$ & & Clustered, ``Volatile", N-body \\
        & & $1.60\:[1.20,\,2.09]$ & $4.50\:[1.69,\,11.5]$ & & Clustered, ``Rocky", N-body \\
        \hline
        \hd{}~f & $22.72 \pm 0.015$ & Unknown & $7.30 \pm 0.40$ & Planet? & \citetalias[][]{Motalebi2015, Gillon2017a}\\
        \hline
        \hd{}~d & $46.86 \pm 0.028$ & Unknown & $16.2 \pm 0.64$ & Planet & \citetalias[][]{Motalebi2015, Gillon2017a}\\
        \hline
        \hd{}~g & $94.2 \pm 0.2$ & Unknown & $11 \pm 1$ & Planet? & \citetalias[][]{Vogt2015}\\
        \hline
        \multirow{8}{*}{H[a] PxP--2} & $174\:[169,\,385]$ & $1.69\:[1.25,\,2.27]$ & $3.99\:[1.95,\,6.36]$ & \multirow{8}{*}{Predicted} & Ratios, ``Volatile", $\Delta = 8$ \\
        & $174\:[168,\,384]$ & $1.79\:[1.33,\,2.41]$ & $6.71\:[2.41,\,18.7]$ & & Ratios, ``Rocky", $\Delta = 8$ \\
        \cline{2-4}
        & $145\:[145,\,399]$ & $1.68\:[1.25,\,2.26]$ & $3.95\:[1.94,\,6.31]$ & & Clustered, ``Volatile", $\Delta = 8$ \\
        & $145\:[145,\,393]$ & $1.77\:[1.32,\,2.39]$ & $6.45\:[2.34,\,18.2]$ & & Clustered, ``Rocky", $\Delta = 8$ \\
        \cline{2-4}
        & $174\:[169,\,365]$ & $1.69\:[1.26,\,2.28]$ & $3.99\:[2.00,\,6.40]$ & & Ratios, ``Volatile", N-body \\
        & $174\:[169,\,364]$ & $1.79\:[1.34,\,2.42]$ & & $6.71\:[2.45,\,18.7]$ & Ratios, ``Rocky", N-body \\
        \cline{2-4}
        & $145\:[145,\,381]$ & $1.68\:[1.26,\,2.27]$ & $3.95\:[1.98,\,6.35]$ & & Clustered, ``Volatile", N-body \\
        & $145\:[145,\,376]$ & $1.77\:[1.33,\,2.41]$ & $6.45\:[2.04,\,6.43]$ & & Clustered, ``Rocky", N-body \\
        \hline
    \end{tabular}
\end{table*}

\begin{table*}[ht]
    \centering
    \begin{tabular}{|l|c|c|c|c|r|}
        \hline
        \multirow{8}{*}{H[b] PxP--2} & $86.3\:[84.8,\,206]$ & $1.65\:[1.23,\,2.23]$ & $3.84\:[1.84,\,6.18]$ & \multirow{8}{*}{Predicted} & Ratios, ``Volatile", $\Delta = 8$ \\
        & $86.3\:[85.4,\,204]$ & $1.72\:[1.28,\,2.31]$ & $5.85\:[2.11,\,16.2]$ & & Ratios, ``Rocky", $\Delta = 8$ \\
        \cline{2-4}
        & $73.9\:[73.9,\,167]$ & $1.65\:[1.22,\,2.22]$ & $3.84\:[1.79,\,6.09]$ & & Clustered, ``Volatile", $\Delta = 8$ \\
        & $74.1\:[74.1,\,169]$ & $1.70\:[1.27,\,2.28]$ & $5.61\:[2.05,\,15.5]$ & & Clustered, ``Rocky", $\Delta = 8$ \\
        \cline{2-4}
        & $86.3\:[84.6,\,205]$ & $1.65\:[1.23,\,2.23]$ & $3.84\:[1.84,\,6.18]$ & & Ratios, ``Volatile", N-body \\
        & $86.3\:[85.2,\,203]$ & $1.72\:[1.28,\,2.31]$ & $5.85\:2.11,\,16.2]$ & & Ratios, ``Rocky", N-body \\
        \cline{2-4}
        & $73.9\:[73.9,\,166]$ & $1.65\:[1.22,\,2.22]$ & $3.84\:[1.79,\,6.09]$ & & Clustered, ``Volatile", N-body \\
        & $74.1\:[74.1,\,168]$ & $1.70\:[1.277,\,2.28]$ & $5.61\:[2.05,\,15.5]$ & & Clustered, ``Rocky", N-body \\
        \hline
        \multirow{8}{*}{H[c] PxP--2} & \multirow{2}{*}{$174\:[169,\,386]$} & $2.13\:[1.30,\,3.81]$ & $5.75\:[2.23,\,14.4]$ & \multirow{8}{*}{Predicted} & Ratios, ``Volatile", $\Delta = 8$ \\
        & & $1.77\:[1.32,\,2.38]$ & $6.45\:[2.34,\,17.9]$ & & Ratios, ``Rocky", $\Delta = 8$ \\
        \cline{2-4}
        & \multirow{2}{*}{$141\:[140,\,389]$} & $1.91\:[1.25,\,3.62]$ & $4.84\:[1.95,\,13.3]$ & & Clustered, ``Volatile", $\Delta = 8$ \\
        & & $1.70\:[1.28,\,2.30]$ & $5.61\:[2.10,\,15.9]$ & & Clustered, ``Rocky", $\Delta = 8$ \\
        \cline{2-4}
        & \multirow{2}{*}{$174\:[169,\,416]$} & $2.14\:[1.30,\,3.84]$ & $5.79\:[2.23,\,14.6]$ & & Ratios, ``Volatile", N-body \\
        & & $1.78\:[1.32,\,2.40]$ & $6.48\:[2.34,\,18.0]$ & & Ratios, ``Rocky", N-body \\
        \cline{2-4}
        & \multirow{2}{*}{$141\:[140,\,414]$} & $1.92,\:[1.25,\,3.63]$ & $4.87\:[1.95,\,13.4]$ & & Clustered, ``Volatile", N-body \\
        & & $1.71\:[1.28,\,2.32]$ & $5.63\:[2.10,\,16.0]$ & & Clustered, ``Rocky", N-body \\
        \hline
        \multirow{8}{*}{H[d] PxP--2} & $86.3\:[85.2,\,206]$ & $2.05\:[1.29,\,4.04]$ & $5.41\:[2.17,\,15.8]$ & \multirow{8}{*}{Predicted} & Ratios, ``Volatile", $\Delta = 8$ \\
        & $86.3\:[85.2,\,204]$ & $1.65\:[1.23,\,2.22]$ & $5.06\:[1.83,\,14.1]$ & & Ratios, ``Rocky", $\Delta = 8$ \\
        \cline{2-4}
        & $70.2\:[69.3,\,80.2]$ & $1.82\:[1.23,\,3.66]$ & $4.48\:[1.84,\,13.5]$ & & Clustered, ``Volatile", $\Delta = 8$ \\
        & $70.3\:[69.3,\,80.2]$ & $1.59\:[1.20,\,2.09]$ & $4.46\:[1.69,\,11.5]$ & & Clustered, ``Rocky", $\Delta = 8$ \\
        \cline{2-4}
        & $86.3\:[84.9,\,205]$ & $2.06\:[1.29,\,4.04]$ & $5.45\:[2.17,\,15.8]$ & & Ratios, ``Volatile", N-body \\
        & $86.3\:[84.9,\,203]$ & $1.66\:[1.23,\,2.23]$ & $5.09\:[1.83,\,14.1]$ & & Ratios, ``Rocky", N-body \\
        \cline{2-4}
        & $70.2\:[69.3,\,80.1]$ & $1.83\:[1.23,\,3.66]$ & $4.52\:[1.84,\,13.5]$ & & Clustered, ``Volatile", N-body \\
        & $70.3\:[69.3,\,80.1]$ & $1.60\:[1.20,\,2.09]$ & $4.49\:[1.69,\,11.5]$ & & Clustered, ``Rocky", N-body \\
        \hline
    \end{tabular}
    \\[10pt]
    \textbf{Notes}: In the PxP naming scheme, H[.] are the different hypotheses as in Section \ref{subsec:hypo}. Radius is measured for planets b and c, predicted from M--R relationship for remaining planets and predictions.  Mass value is given for transit-verified planets b and c as well as predicted planets, whereas m $\sin$ i value is given for RV-verified planets d, f, and g with unknown inclination.  Planets denoted ``Planet?" are the two planets that we tested removing from the system.  The specific \textsc{DYNAMITE} period model, mass-radius relation, and dynamical stability criterion used are noted in the last column for the predicted planets.  The inclination and eccentricity models were fixed as the multiplicity-dependent Lognormals from SysSim \citep[][]{He2020}.
\end{table*}

\begin{table*}[ht]
    \centering
    \caption{\hd{} Planet and Planet Candidate Inclination and Eccentricity}
    \label{tab:planets2}
    \begin{tabular}{|l|c|c|c|r|}
        \hline
        \textbf{Name} & \textbf{Inclination} & \textbf{Eccentricity} & \textbf{Note} & \textbf{Origin/Reference}\\
        \hline
        \hd{}~b & $85.05 \pm 0.09$ & 0 & Planet & \citetalias[][]{Motalebi2015, Gillon2017a}\\
        \hline
        \hd{}~c & $87.28 \pm 0.10$ & $0.062 \pm 0.039$ & Planet & \citetalias[][]{Motalebi2015, Gillon2017a}\\
        \hline
        \hd{}~f & Unknown & $0.148 \pm 0.047$ & Planet? & \citetalias[][]{Motalebi2015, Gillon2017a}\\
        \hline
        \hd{}~d & Unknown & $0.138 \pm 0.025$ & Planet & \citetalias[][]{Motalebi2015, Gillon2017a}\\
        \hline
        \hd{}~g & Unknown & 0 & Planet? & \citetalias[][]{Vogt2015}\\
        \hline
        \multirow{8}{*}{H[a] PxPs} & \multirow{2}{*}{$82.0\:[79.1,\,85.1]$} & $0.024\:[0.009,\,0.036]$ & \multirow{8}{*}{Predicted} & Rayleigh+Iso, Rayleigh, $\Delta = 8$ \\
        \cline{3-3}
        & & $0.023\:[0.012,\,0.045]$ & & Rayleigh+Iso, Lognormal, $\Delta = 8$ \\
        \cline{2-3}
        & \multirow{2}{*}{$85.6\:[84.3,\,86.7]$} & $0.024\:[0.009,\,0.036]$ & & Lognormal, Rayleigh, $\Delta = 8$ \\
        \cline{3-3}
        & & $0.023\:[0.012,\,0.045]$ & & Lognormal, Lognormal, $\Delta = 8$ \\
        \cline{2-3}
        & \multirow{2}{*}{$82.0\:[79.2,\,85.0]$} & $0.024\:[0.009,\,0.036]$ & & Rayleigh+Iso, Rayleigh, N-body \\
        \cline{3-3}
        & & $0.023\:[0.012,\,0.045]$ & & Rayleigh+Iso, Lognormal, N-body \\
        \cline{2-3}
        & \multirow{2}{*}{$85.6\:[84.4,\,86.6]$} & $0.024\:[0.009,\,0.036]$ & & Lognormal, Rayleigh, N-body \\
        \cline{3-3}
        & & $0.023\:[0.012,\,0.045]$ & & Lognormal, Lognormal, N-body \\
        \hline
        \multirow{8}{*}{H[b] PxPs} & \multirow{2}{*}{$82.0\:[79.1,\,85.1]$} & $0.024\:[0.012,\,0.039]$ & \multirow{8}{*}{Predicted} & Rayleigh+Iso, Rayleigh, $\Delta = 8$ \\
        \cline{3-3}
        & & $0.031\:[0.016,\,0.063]$ & & Rayleigh+Iso, Lognormal, $\Delta = 8$ \\
        \cline{2-3}
        & \multirow{2}{*}{$85.6\:[84.1,\,87.3]$} & $0.024\:[0.012,\,0.039]$ & & Lognormal, Rayleigh, $\Delta = 8$ \\
        \cline{3-3}
        & & $0.031\:[0.016,\,0.063]$ & & Lognormal, Lognormal, $\Delta = 8$ \\
        \cline{2-3}
        & \multirow{2}{*}{$82.0\:[79.2,\,85.1]$} & $0.024\:[0.012,\,0.039]$ & & Rayleigh+Iso, Rayleigh, N-body \\
        \cline{3-3}
        & & $0.031\:[0.016,\,0.063]$ & & Rayleigh+Iso, Lognormal, N-body \\
        \cline{2-3}
        & \multirow{2}{*}{$85.6\:[84.0,\,87.3]$} & $0.024\:[0.012,\,0.039]$ & & Lognormal, Rayleigh, N-body \\
        \cline{3-3}
        & & $0.031\:[0.016,\,0.063]$ & & Lognormal, Lognormal, N-body \\
        \hline
        \multirow{8}{*}{H[c] PxPs} & \multirow{2}{*}{$82.0\:[79.1,\,85.1]$} & $0.024\:[0.012,\,0.039]$ & \multirow{8}{*}{Predicted} & Rayleigh+Iso, Rayleigh, $\Delta = 8$ \\
        \cline{3-3}
        & & $0.031\:[0.016,\,0.063]$ & & Rayleigh+Iso, Lognormal, $\Delta = 8$ \\
        \cline{2-3}
        & \multirow{2}{*}{$85.6\:[84.1,\,87.3]$} & $0.024\:[0.012,\,0.039]$ & & Lognormal, Rayleigh, $\Delta = 8$ \\
        \cline{3-3}
        & & $0.031\:[0.016,\,0.063]$ & & Lognormal, Lognormal, $\Delta = 8$ \\
        \cline{2-3}
        & \multirow{2}{*}{$82.0\:[79.1,\,85.0]$} & $0.024\:[0.012,\,0.039]$ & & Rayleigh+Iso, Rayleigh, N-body \\
        \cline{3-3}
        & & $0.031\:[0.016,\,0.063]$ & & Rayleigh+Iso, Lognormal, N-body \\
        \cline{2-3}
        & \multirow{2}{*}{$85.6\:[84.1,\,87.2]$} & $0.024\:[0.012,\,0.039]$ & & Lognormal, Rayleigh, N-body \\
        \cline{3-3}
        & & $0.031\:[0.016,\,0.063]$ & & Lognormal, Lognormal, N-body \\
        \hline
        \multirow{8}{*}{H[d] PxPs} & \multirow{2}{*}{$82.7\:[78.3,\,86.7]$} & $0.024\:[0.012,\,0.039]$ & \multirow{8}{*}{Predicted} & Rayleigh+Iso, Rayleigh, $\Delta = 8$ \\
        \cline{3-3}
        & & $0.046\:[0.023,\,0.092]$ & & Rayleigh+Iso, Lognormal, $\Delta = 8$ \\
        \cline{2-3}
        & \multirow{2}{*}{$86.1\:[83.8,\,88.6]$} & $0.024\:[0.012,\,0.039]$ & & Lognormal, Rayleigh, $\Delta = 8$ \\
        \cline{3-3}
        & & $0.046\:[0.023,\,0.092]$ & & Lognormal, Lognormal, $\Delta = 8$ \\
        \cline{2-3}
        & \multirow{2}{*}{$82.7\:[78.3,\,86.7]$} & $0.024\:[0.012,\,0.039]$ & & Rayleigh+Iso, Rayleigh, N-body \\
        \cline{3-3}
        & & $0.046\:[0.023,\,0.092]$ & & Rayleigh+Iso, Lognormal, N-body \\
        \cline{2-3}
        & \multirow{2}{*}{$86.1\:[83.8,\,88.6]$} & $0.024\:[0.012,\,0.039]$ & & Lognormal, Rayleigh, N-body \\
        \cline{3-3}
        & & $0.046\:[0.023,\,0.092]$ & & Lognormal, Lognormal, N-body \\
        \hline
    \end{tabular}
    \\[10pt]
    \textbf{Notes}: In the PxP naming scheme, H[.] are the different hypotheses for the existence of the known system planets as in Section \ref{subsec:hypo}.  \hd{}~b likely has a tidally circularized orbit due to its semi-major axis, so its eccentricity is set to 0 \citepalias[][]{Gillon2017a}, and the eccentricity of planet g was set to 0 by \citetalias[][]{Vogt2015}.  Planets denoted ``Planet?" are the two planets that we tested removing from the system in the different hypotheses.  The specific \tnt{} inclination and eccentricity models and dynamical stability criterion used are noted in the last column for the predicted planets.  The orbital period model and mass-radius relationship were fixed at the period ratios from EPOS \citep[][]{Mulders2018} and the ``volatile-rich" M-R power-law \citep[][]{Otegi2020}
\end{table*}

\subsection{Inclination and eccentricity}

\begin{figure*}
    \centering
    \includegraphics[width=1.05\columnwidth]{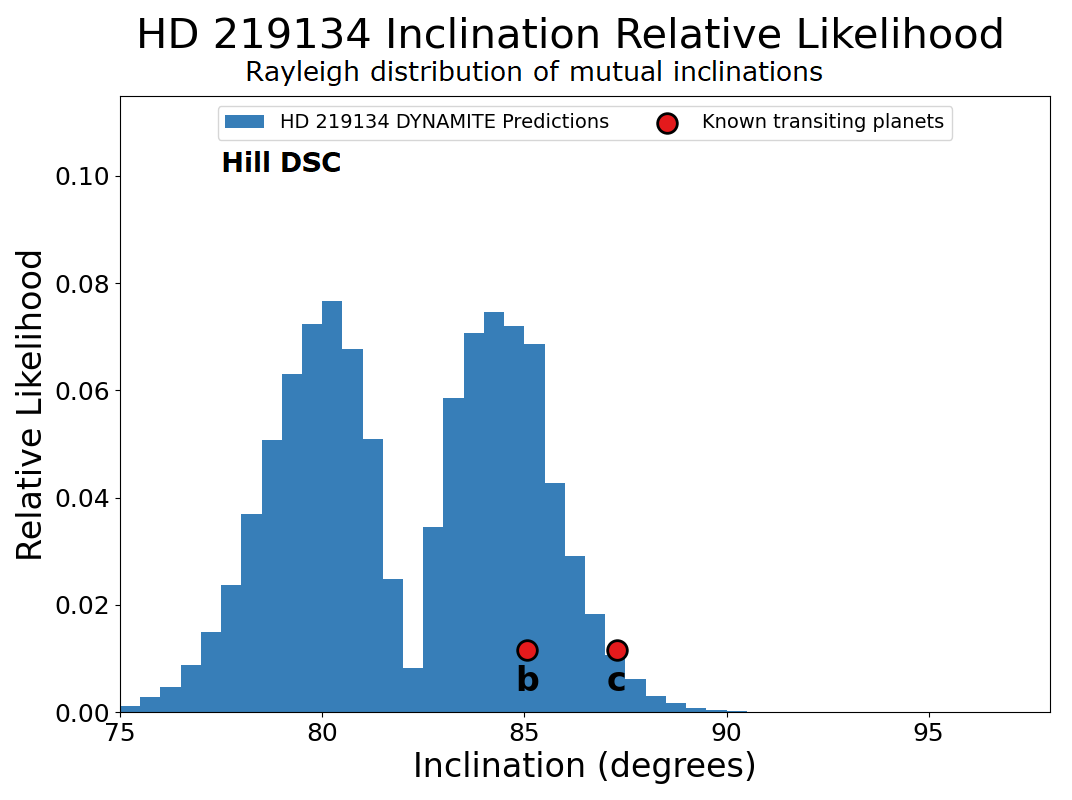}
    \includegraphics[width=1.05\columnwidth]{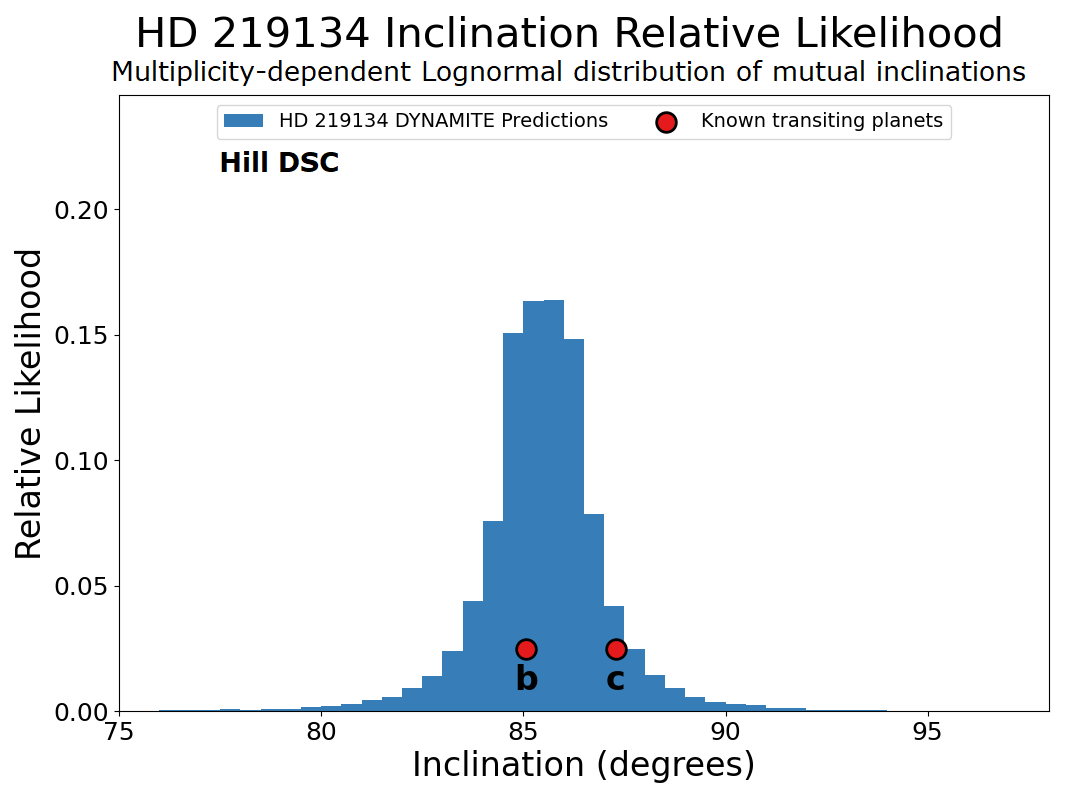}\\
    \vspace{10pt}
    \includegraphics[width=1.05\columnwidth]{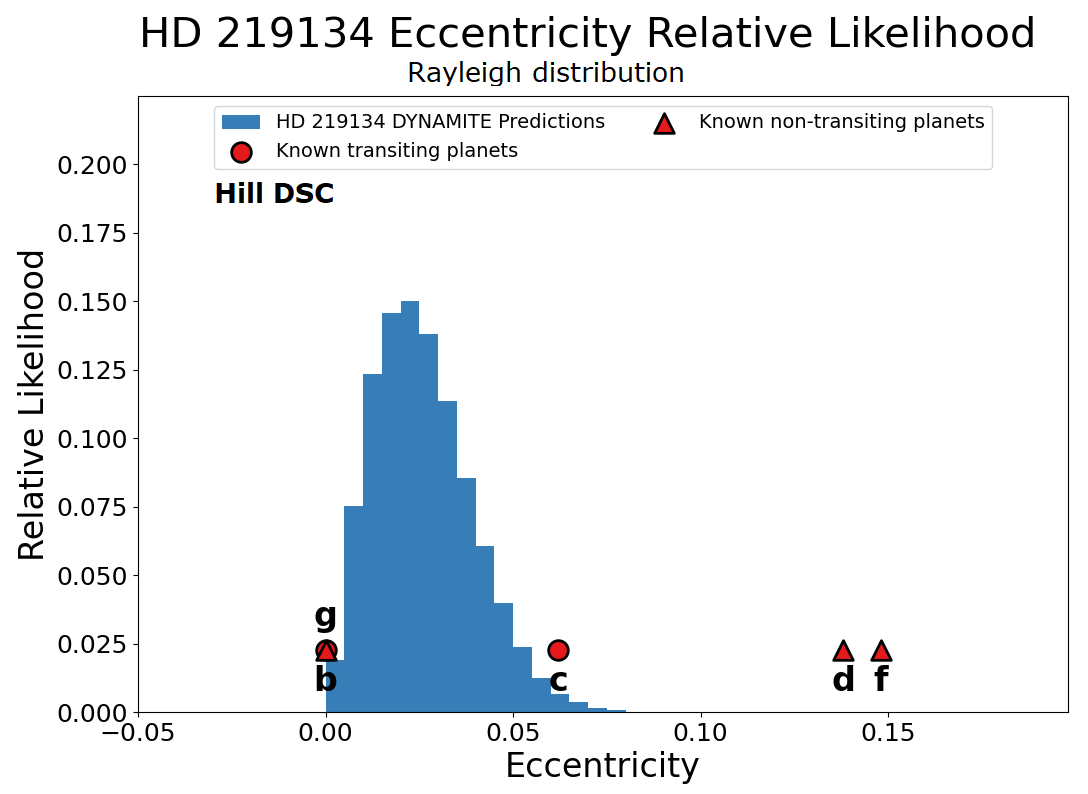}
    \includegraphics[width=1.05\columnwidth]{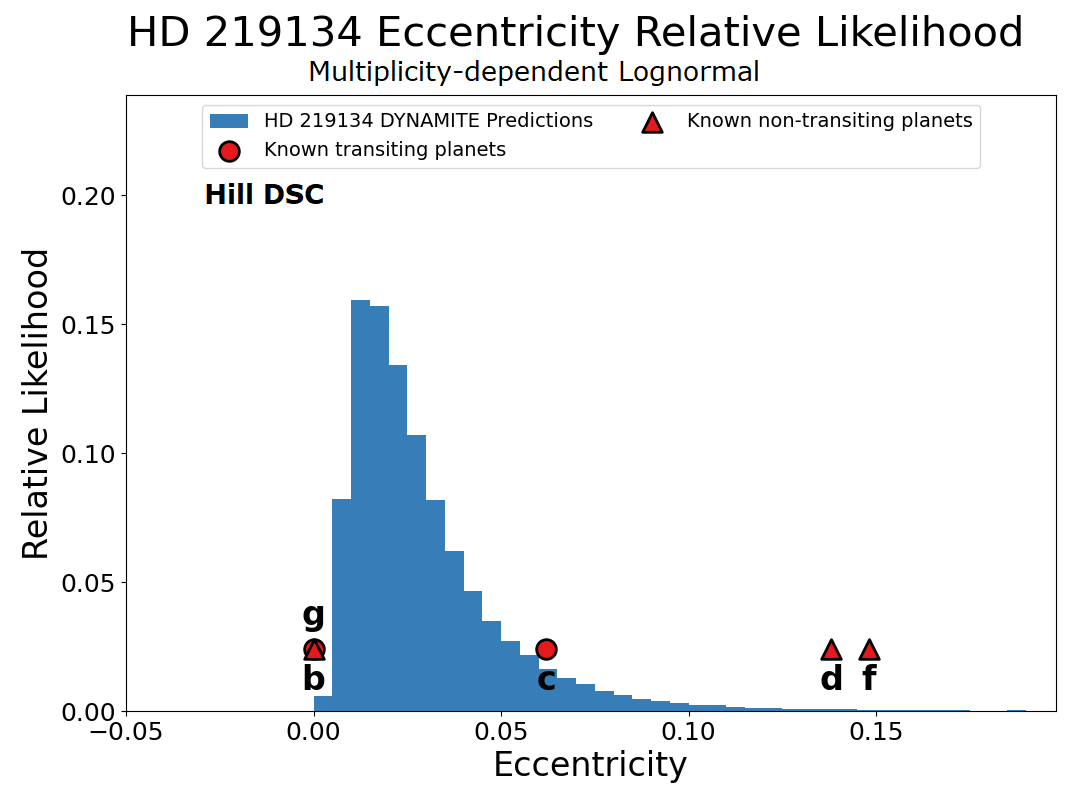}
    \caption{The histograms for the inclinations (top) and eccentricities (bottom) of the predicted injections from \tnt{} for Hypothesis H[a]; the inclinations for planets b-d are unknown and are therefore not known to the \tnt{} algorithm nor included in the analysis.  The left figures use the Rayleigh + isotropic fraction model for inclinations and Rayleigh model for eccentricities, while the right ones use the multiplicity-dependent Lognormals from the analysis by \citet[][]{He2020}.  All of the above are using the equal ratios period model, the ``volatile-rich" mass-radius relation, and the mutual Hill radius dynamical stability criterion.}
    \label{fig:inc_ecc}
\end{figure*}

We added a new inclination model to test the predictions between them, and also incorporated two different eccentricity models for the dynamical stability analysis that required eccentricities in the N-body integrations.  The results from the analysis on the \hd{} system using the different statistical models are shown in Figure~\ref{fig:inc_ecc}.  For these results, the other population modules were held constant (equal period ratios, ``volatile-rich" M--R relation, and mutual Hill radius stability criterion) to highlight the specific effect of the inclination and eccentricity choices.

The original inclination model, which fits the known mutual inclinations of the system to a Rayleigh distribution with parameter $2^\circ$ around a central system inclination, found that the transiting planets were likely to be on the ``high" side of this mutual inclination distribution, since the remaining planets have not been known to transit.  Thus, the system inclination was fit to $\sim\:82^\circ$ (where $0^\circ$ is face-on in the plane of the sky and $90^\circ$ is edge-on), with the inclination likelihood therefore peaking at $80^\circ$ and $84^\circ$.  The new inclination model, which fits the known mutual inclinations of the system to a multiplicity-dependent Lognormal distribution, found unsurprisingly that the system inclination was likely to be near the inclination of the two known transiting planets.  This is still consistent with the likelihood of the other known planets not transiting, as the limiting inclination for transits approaches $90^\circ$ as the orbital period increases.  Until additional information is gathered from the system, it is unlikely we will be able to further constrain these parameters.

For the simple model of planet eccentricities, there is no difference between any of the system architecture hypotheses, and so the predictions follow the Rayleigh distribution, with some slight deviations likely due to injections that are rejected due to other parameters.  For the multiplicity-dependent Lognormal, we find that higher eccentricities can be expected in general than the Rayleigh distribution, and the mean of the distribution is higher in Hypothesis H[d] ($\mu = 0.046$) than in Hypothesis H[a] ($\mu = 0.023$).  However, none of the models show a high likelihood for having one or multiple planets with moderate eccentricities, as seen in the \hd{} system.

\subsection{Additional planets}

\begin{figure}[t]
    \centering
    \includegraphics[width=\columnwidth]{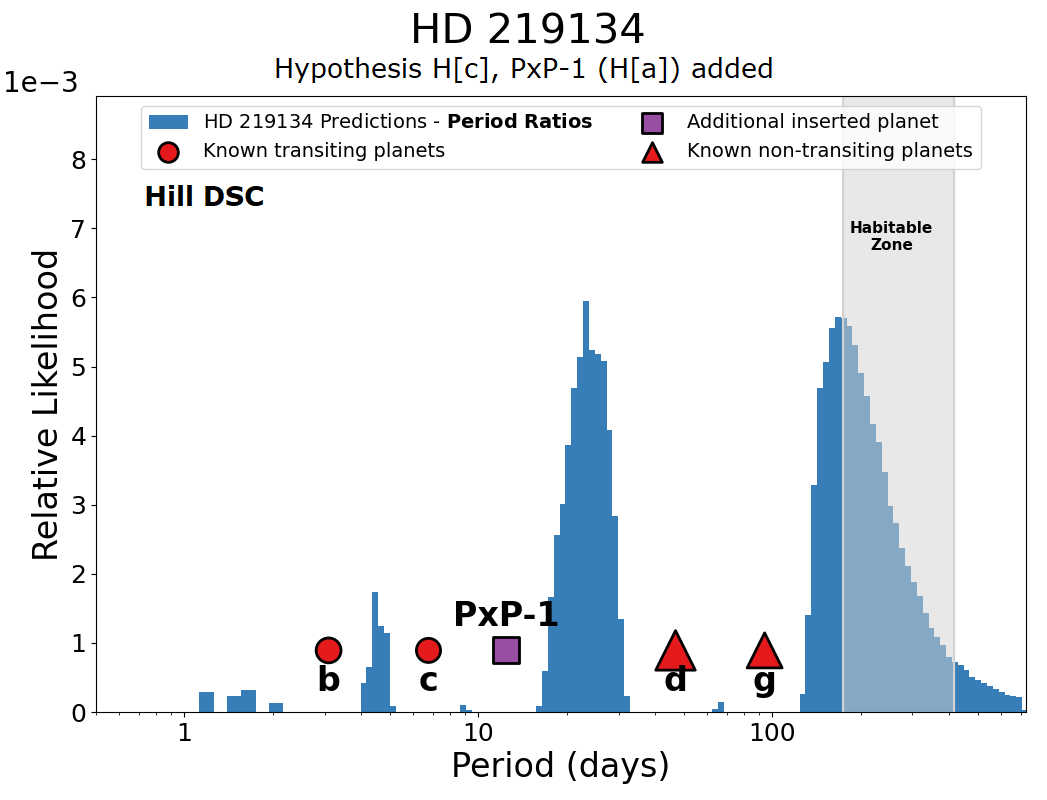}
    \caption{The relative likelihood assuming Hypothesis H[c] for the known planets, but the position of PxP--1 comes from the peak in relative likelihood from Hypothesis H[a].  This shows that the lower and more spread out likelihood of injections seen in Hypotheses H[c] and H[d] is indicative of two potential planets.}
    \label{fig:add12}
\end{figure}

\textbf{Hypothesis H[a]}: We find two peaks in the orbital period relative likelihood at $\sim$12 days (PxP--1 H[a]) and at $\sim$174 days (PxP--2 H[a]).  The first peak lies between planet c and planet candidate f, and it can be used to determine the likelihood of planet candidate f being a genuine planet in Hypotheses H[c] and/or H[d] (see Figure~\ref{fig:add12}).  The second peak lies beyond planet candidate g and inside the habitable zone of \hd{}, near the inner edge \citep[e.g.,][]{Kopparapu2013}.  However, as this second peak is an extrapolation beyond the last available data point, the predictions are less constraining in the space beyond the outermost known planet.

\textbf{Hypothesis H[b]}: We again find two peaks in the orbital period likelihood.  The first at 12 days corresponds with PxP--1 H[a], which we now call PxP--1 (H[a], H[b]).  The second peak, however, moves inward from the peak in Hypothesis H[a], to $\sim$ 90 days.  Because planet candidate g has been excluded from the analysis, we now find the peak for PxP--2 H[b] likely corresponds to planet candidate g.

\textbf{Hypothesis H[c]}: Here, the peak around 12 days in the previous hypotheses has been flattened and widened out with the removal of planet candidate f from the analysis.  Similar behavior was seen in our past studies \citep[][]{Dietrich2020, Dietrich2021}, where we found evidence that this shape indicates that not one, but two planets are present.  When PxP--1 (H[a], H[b]) is added to Hypothesis H[c], we do recover a sharp peak very near the orbital period of planet candidate f, as in Figure~\ref{fig:add12}.  Thus, we find this gap could likely hide two planets.  The outer peak from Hypothesis H[a] we denoted as PxP--2 H[a] is also visible here, so we update its classification to PxP--2 (H[a], H[c]).

\textbf{Hypothesis H[d]}: We see here the same spread-out pattern between 10 and 30 days indicative of two planets from Hypothesis H[c].  We also find the outer peak has again shifted as in Hypothesis H[b] to a similar position as PxP--2 H[b], so PxP--2 (H[b], H[d]) does provide support for planet candidate g.

\subsection{N-body integrations}

Our results from the dynamical stability criteria that utilized N-body integrations differed for each of the planet hypotheses.  The spectral fraction analysis showed very similar results to the mutual Hill radius, even in areas near the mutual Hill radius boundary of 8 (see Figure~\ref{fig:dsc_test1}).  There were some slight significant differences, including in areas near orbital period resonances, but most of the difference was found to be random noise generated from the Monte Carlo iterations.

SPOCK's feature classifier finds that the system architectures for Hypotheses H[a]--H[d] would be stable without injecting planets, as the simple stability predictor returns $0.479 \pm 0.028$, $0.648 \pm 0.036$, $0.636 \pm 0.022$, and $0.651 \pm 0.023$, respectively, for each of the hypotheses.  However, when additional planets are injected via Monte Carlo methods, the system under Hypothesis H[a] is pushed just past the stability limit of 0.34 from predictor (set by the 10\% false positive rate), down to $0.328 \pm 0.092$.  The remaining Hypotheses H[b]--H[d] have simple stability predictors of $0.868 \pm 0.116$, $0.510 \pm 0.047$, and $0.910 \pm 0.052$, which are all considered stable.

SPOCK's deep regressor also shows that all four hypotheses as they stand are stable, with the stability predictor having values of $0.407 \pm 0.021$, $0.520 \pm 0.030$, $0.635 \pm 0.020$, and $0.824 \pm 0.019$.  Adding in another planet to Hypothesis H[a], however, causes the stability predictor to drop to $0.135 \pm 0.025$, and no additional planet was considered stable long-term.  In addition, injecting a planet in Hypothesis H[b] also drops the stability predictor down to $0.246 \pm 0.046$, again with no additional planet found to be stable.  For Hypothesis H[c], the stability predictor was $0.349 \pm 0.059$ over the injection tests, with 60\% of the injections accepted as long-term stable.  Hypothesis H[d] was stable with respect to injections with a stability predictor value of $0.473 \pm 0.080$.

\begin{figure*}[ht]
    \centering
    \includegraphics[width=1.58\columnwidth]{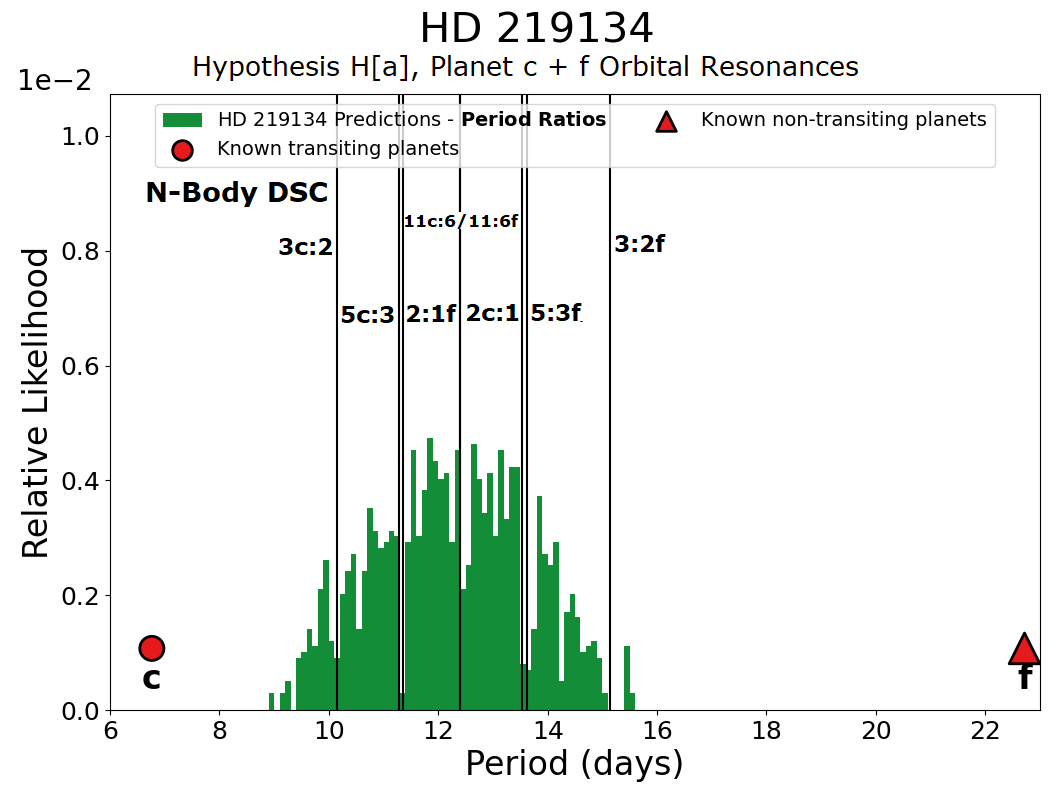}
    \includegraphics[width=1.58\columnwidth]{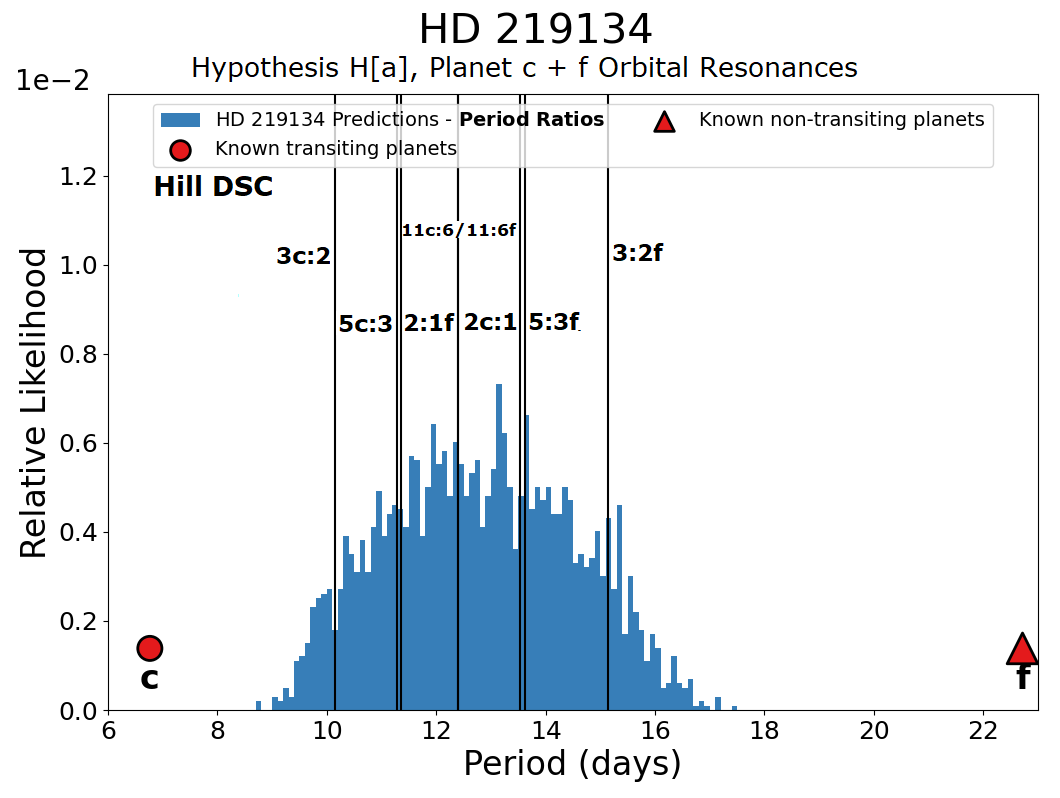}
    \caption{The 2:1, 3:2, 5:3, and 11:6 resonances plotted for an additional planet along with planets c and f in Hypothesis H[a].  2:1 and 3:2 were chosen as low-order strong resonances, whereas 5:3 and 11:6 are shown because they overlapped a 2:1 resonance or the 11:6 resonance of the other known planet.  Top: Injections (in green) via the $\sim10^6$-orbit N-body integration dynamical stability criterion, which show lower probabilities of injection near the resonances.  Bottom: Injections (in blue) via the mutual Hill radius dynamical stability criterion, which shows no such consistent effect.  The difference in y-scale is due to finding fewer planets overall in this gap, as the likelihood gets normalized to 1 across all injections.}
    \label{fig:resonances}
\end{figure*}

The N-body integrations did find dips in the relative likelihood near resonances, a feature that is missed by the mutual Hill radius criterion, which is focused more on the space closer to planets.  Figure~\ref{fig:resonances} shows the difference between the spectral fraction analysis and the mutual Hill radius for the predicted planets between planets c and f in Hypothesis H[a].  The low-order resonances 2:1 and 3:2 were located at dips in the injections in period space, and additional higher-order resonances 5:3 and 11:6 were added as they coincided with either the same resonance for both planets (11:6) or the 2:1 resonance for the other planet (5:3).  The mutual Hill radius criterion does not see this effect.

\section{Discussion -- \tnt{} Predictions} \label{sec:TNT_disc}

\subsection{Inclination and eccentricity distributions}

For planets without inclination measurements and $m \sin i$ values measured by radial velocity observations, the true masses can vary widely due to the assumption of system inclination isotropy, which widens the resulting predicted distribution for the predicted planet mass.  For isotropically-distributed inclinations, the potentially large difference in inclination between neighboring planets also affects their dynamical stability in the N-body integrations, decreasing the available space for injecting stable planets between them.  Allowing for unknown inclinations provides the most accurate prediction of the likelihood for planet injections, especially when utilizing N-body integrations.

Introducing the eccentricity to the planet parameters for which \tnt{} tracks in its analysis provides an additional measure in the dynamical stability.  While the mutual Hill radius criterion is independent of the eccentricity of the planets, it is easy to determine that a highly eccentric planet would strongly affect the dynamical stability in a way that the mutual Hill radius criterion would be blind to.  While it is somewhat unlikely to have highly eccentric planets in higher-multiplicity systems \citep[e.g.,][]{He2020}, our ability to include the eccentricities of the known planets in our predictions helps increase the accuracy of our dynamical stability models and constrain the available parameter space for additional planets.

\subsection{Support for planets f and g}

\textbf{Planet f}

Based on their RV measurements, \citetalias[][]{Vogt2015} detailed the presence of planet f at 22.81 days (not found in the analysis by \citetalias[][]{Motalebi2015}), and noted the closeness to the rotational period gathered from $v \sin i$ measurements.  By testing different segments of the radial velocity curve for rotational spot modulation residual signals, \citetalias[][]{Vogt2015} determined the 22.8-day signal was likely to be real and not a result of the stellar rotation.  However, a follow-up analysis by \citet[][]{Johnson2016} narrowed in stellar activity indicators to a period of $22.83 \pm 0.03$ days, very similar to the reported period of planet f.  Thus, planet f is reported as ``controversial" in the NASA Exoplanet Archive, even though a follow-up analysis of the transiting planets by \citetalias[][]{Gillon2017a} includes planet f in its study.

\textbf{Planet g}

The analysis by \citetalias[][]{Vogt2015} also announced the presence of planet g at 94.2 days.  This planet was also not found in the fit by \citetalias[][]{Motalebi2015}, and was not even mentioned in the analysis by \citetalias[][]{Gillon2017a}.  \citet[][]{Johnson2016} mentioned it in a passing reference to the long rotational period measured via $v \sin i$ by \citetalias[][]{Motalebi2015}, but do not provide any insight into the planet's existence nor re-assess the data gathered on the planet.  Thus, the only data gathered for this planet come from the \citetalias[][]{Vogt2015} study, and the uncertainties on the measured parameters are much larger than the other planets.  Still, planet g does not hold the ``controversial" status that planet f has in the Exoplanet Archive, even though its existence is largely centered on only one analysis.

\textbf{\tnt{} analysis}

To assess the likelihood of these planets being found at their current positions, we show that if \hd{} f is excluded and a planet is to be added interior to the orbit of planet d at 46 days, the period of planet f is very near one of two local maxima of the relative likelihood in period space.  With no planet f, the relative likelihood is spread out between 6 and 46 days.  However, if planet f is added, then PxP-1 is predicted around 12 days, and conversely if PxP-1 is then added but planet f is excluded, we find an additional predicted planet that matches the period of planet f.  In addition, if \hd{} g is excluded and a planet is to be added exterior to the orbit of planet d, the period of planet g is very close to the local maximum there as well.  Both of these findings are true for the mutual Hill radius cutoff and the N-body integrations; the main difference between them is that the local maximum for planet g is pushed slightly further out for the spectral fraction analysis, likely due to the moderately high eccentricity of planet d.

\subsection{Predicted planets in the HD 219134 system} \label{subsec:add}

Depending on the choice of dynamical stability criterion, we predict 1-3 additional planets in the system. We find that an additional planet we label PxP--1 can exist between planets c and f, with an orbital period of $\sim$12 days.  If planet f is excluded, we find that this single planet likelihood gets stretched in period space and lowered in probability, which we have found is a good indicator that multiple planets can be injected in the gap iteratively \citep[for more details see][]{Dietrich2020}.  Therefore, we find a high probability of recovering planet f and still adding PxP--1 around 12 days.  This is true for the simple mutual Hill radius dynamical stability criterion as well as both the N-body integration-based SPOCK deep regression and spectral fraction analyses of the \hd{} system under each system architecture hypothesis.

\begin{figure}[t]
    \centering
    \includegraphics[width=\columnwidth]{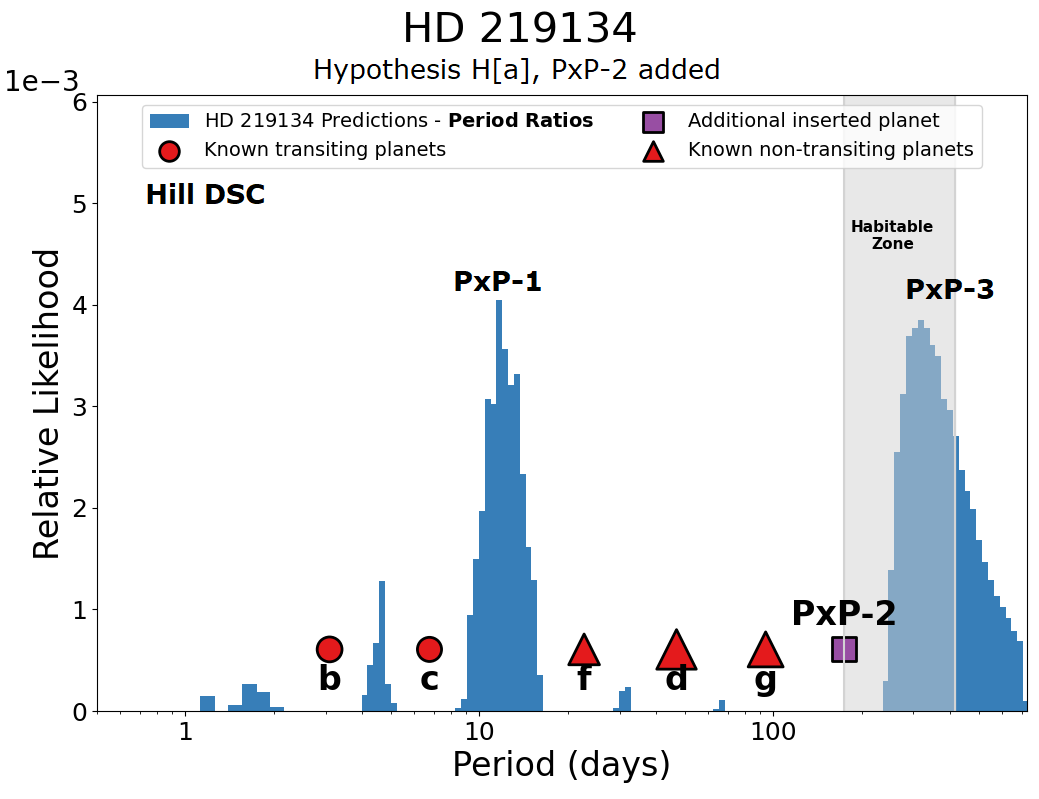}
    \caption{The relative likelihood assuming Hypothesis H[a] for the known planets, but adding in predicted planet PxP--2 also from Hypothesis H[a], utilizing the mutual Hill radius dynamical stability criterion.  This shows an additional predicted planet at around 320 days, in the outer part of the habitable zone.}
    \label{fig:add21}
\end{figure}

If planet g is excluded from the system we recover it well, and with it in the system we also predict an additional planet PxP--2 with an orbital period of $\sim$174 days.  This places PxP--2 at the inner edge of the likely habitable zone for \hd{}.  If the relatively equal period ratios through the planets (including PxP--1) continues beyond planet g through PxP--2, another hypothetical planet could exist with an orbital period of $\sim$321 days, in the outer part of the habitable zone (see Figure~\ref{fig:add21}).  This remains true for all of the implementations of the dynamical stability criterion, as the given and expected eccentricities of planet g and PxP--2 are low, which lessens any effect of the difference between the implementations.  We find that an additional planet of $\sim4\:M_\oplus$ (near the median of the predicted planet mass distribution) could have an orbital period as close as $\sim$ 140 days and still be stable with respect to planet g, whereas a planet of $\sim11\:M_\oplus$ (similar to the masses of the two most exterior planets) could only be located exterior to 150 days and still be stable with respect to planet g.

\citet[][]{Rosenthal2021} reported three false positive radial velocity signals in the \hd{} system that they attribute to annual and/or instrumental systematics.  One of these signals has a period of $\sim$28 years, which is close to the entire length of their legacy survey.  However, the other two signals are at $192.06^{+0.40}_{-0.49}$ days and $364.3^{+1.9}_{-2.3}$ days.  While it is likely that these are false positives from an annual and semi-annual systematic from Earth-based measurements, these signals are relatively close to where we would expect additional planets to exist in this system.  The reported respective semi-amplitudes of $2.0^{+0.21}_{-0.20}$ m/s and $1.66^{+0.35}_{-0.31}$ m/s would correspond to $m \sin i$ values of $13.0^{+2.74}_{-2.43}\:M_{\oplus}$ and $15.7^{+1.64}_{-1.57}\:M_{\oplus}$, which would lie in between the reported values for planets d and g, the two most massive planets in the inner system and the two furthest out (disregarding the gas giant in the outer system).  These values would either be outside or just inside the 16th-84th percentile range for the predicted mass using the volatile-rich mass-radius relationship (depending on which hypothesis of the system architecture is utilized for the prediction), but would likely still lie within that range for the rocky mass-radius relationship for all system architectures.

\section{Discussion -- Dynamical Stability} \label{sec:stable_disc}

Our new dynamical stability criteria utilize N-body integrations to determine the level of stability in a system to a much higher degree of accuracy.  However, as using N-body integrations to the standard of $10^9$ orbits of the innermost planet is extremely computationally expensive, both of our new models run for shorter integrations ($10^4$ orbits for SPOCK and $10^6$ orbits for the spectral fraction analysis) and predict the likelihood of stability extended out to $10^9$ orbits from the results of the short integrations.  Whereas the mutual Hill radius measure is essentially instantaneous and thousands of Monte Carlo realizations can be run per second on a standard laptop without utilizing multi-process threads, the N-body integrations need thousands of CPU hours on multiple nodes of a high-performance computer cluster to be run efficiently.

\begin{figure*}
    \centering
    \includegraphics[width=2\columnwidth]{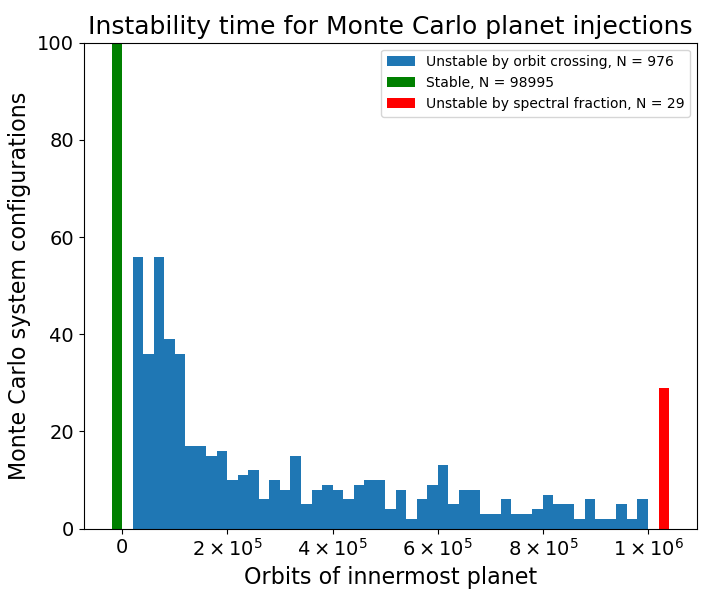}
    \caption{The number of orbits until instability in the N-body integrations for the $10^5$ Monte Carlo realizations of the system under Hypothesis H[a].  Almost 99\% of the injections were considered stable (green), whereas 29 total integrations were found unstable via spectral fraction analysis after surviving the N-body integration (red).}
    \label{fig:tunstable}
\end{figure*}

For the N-body analyses, we count Monte Carlo realizations of systems with injections as unstable if they do not survive the initial integration times.  SPOCK simply returns a boolean value for the system survival (i.e., without hyperbolic orbits) in the first $10^4$ orbits of the innermost planet, but for the spectral fraction analysis we manually reject crossing orbits as well as ejected planets.  Figure~\ref{fig:tunstable} shows the integration times at which different Monte Carlo realizations of \hd{} under Hypothesis H[a] went unstable in the N-body integration, as well as showing the spectral fraction analysis for those that survived the initial $10^6$ orbits of the innermost planet.

Not all orbital crossings result in prompt planetary collisions and/or ejections (e.g., Neptune and Pluto's orbits cross).  In particular, pairs of similar-mass planets can actually persist in such an unstable configuration for tens to hundreds of Myr timescales if the mutual inclinations between planets are large enough \citep[$\gtrsim 1^\circ$;][]{Rice2018}.  However, for the \hd{} system, the relatively large eccentricities and (likely) mutual inclinations of order $\gtrsim1^\circ$ would likely cause increasing instability of crossing orbits, and for the $\sim$10 Gyr age of this system we would expect most crossing orbits to become unstable on this timescale.

\subsection{Mutual Hill radius}

The mutual Hill radius measure of dynamical closeness has been a predictor of stability for decades \citep[see e.g.,][]{Gladman1993}.  If a pair of neighboring planets' semimajor axes are within a certain distance of each other, based on their mass and measured by their Hill spheres of influence, then they will eventually cause the planets to become unstable.  This measure provides a simple and fast prediction for any pair of neighboring planets, as the only calculations it requires are basic arithmetic.

The theoretical limit for stability of two-planet systems is that the difference of their semimajor axes (in units of their mutual Hill radius) should not be smaller than $\Delta = 2 \sqrt 3 \approx 3.46$.  However, there tends to be a distribution in the mutual Hill radius for exoplanet pairs instead of a sharp cutoff \citep[e.g.,][]{Malhotra2015}.  In practice, it is found that systems of three or more planets require significantly larger orbital separation of neighboring planets; \citet[][]{Pu2015} reported $\Delta \gtrsim 12$ for long-term stability of Kepler multis, while \citet[][]{Gilbert2020} report that a large percentage of known exoplanet systems cluster around $\Delta = 20$.  In the present study, we adopted $\Delta = 8$ as a minimum separation, following \citet[][]{He2019}.  However, there are systems this assessment would classify as unstable that are long-lived due to other considerations \citep[e.g., Kepler-36;][]{Carter2012}, so there are cases when this measure is insufficient.  For the \hd{} system across all of the system architecture hypotheses, we find very little difference between the mutual Hill radius and the other configurations; $\sim 1.1\%$ of injections were found to be unstable via the Hill radius as compared to $\sim 1.5\%$ for the spectral fraction analysis.

\subsection{SPOCK}

The utilization of machine learning and training on hundreds of thousands of planetary systems provides a marked improvement in speed over full N-body integrations as well as accuracy over the simpler near-instantaneous methods.  SPOCK's simple feature classifier and deep regressor both perform much faster than the ground-truth full N-body simulations on which they are trained, as well as the spectral fraction analysis, which is performed on integrations 100x longer than SPOCK.  SPOCK provides a probabilistic measure of stability for a single system architecture based on the data sets from which it has learned.  However, while the training sets attempt to cover as many system architectures as possible, they can cover only a small subset of the uncountable infinity of possible planetary architectures, and therefore it may not provide as accurate of a stability analysis that a full N-body integration could provide.

In our study, SPOCK shows a high level of instability for the more complicated system architecture hypotheses.  SPOCK's FeatureClassifier find s that hypotheses including planet g are shown to be less stable than those where it is excluded, as Hypotheses H[a] and H[c] have stability predictors $\lesssim 0.5$ while H[b] and H[d] are $\gtrsim0.85$.  This suggests that the presence of planet g (and any additional planets beyond it) might negatively affect the stability of the other planets interior to planet g.  In addition, the hypotheses including planet f tend to have a higher scatter in the likelihood of stability, as H[a] and H[b] have stability predictor standard deviations of 0.092 and 0.116 as compared to 0.047 and 0.052 for H[c] and H[d].  This likely is a consequence of the moderate reported eccentricity of planet f and the relatively high probability of injecting a planet near 12 days, between planets c and f.

SPOCK's DeepRegressor provides additional samples of the short N-body integration to determine the probability of stability at a more in-depth level while sacrificing some of its speed.  The DeepRegressor showed similar results, with all four system architecture hypotheses stable without any additional planets, albeit with a lower stability predictor than the corresponding FeatureClassifier value.  However, adding in another planet to both Hypotheses H[a] and H[b] causes their stability predictors to drop well below the stability threshold, while Hypothesis H[c] remains barely above the value.  Hypothesis H[d] is still stable with injections, but its stability predictor is lower than all of the values without additional planets except for the five-planet architecture in Hypothesis H[a].  In other words, the DeepRegressor finds the \hd{} system is relatively near the stability limit even with only the three confirmed planets b-d.

In general, if test planets are injected into Hypothesis H[a] anywhere, or into Hypotheses H[b]--H[d] at places that do not specifically correspond to the period of the planet candidates excluded from the analysis of the system, then SPOCK finds the injections would likely cause the system to become unstable (and are therefore rejected).  The SPOCK analysis differs from the mutual Hill radius measure and the spectral fraction analysis of \hd{} in this way, as the other two do not have as large a fraction of rejected injections.  In general, SPOCK predicts that the currently known 3-5 planets in the \hd{} system with orbital periods within 100 days would place the system on the brink of long-term orbital dynamical stability, with little to no room for additional planets.

\subsection{Spectral fraction}

The spectral fraction analysis is closer to the ``ground-truth" of the full N-body integrations with longer test runs, but still provides a usable speed for a system given enough computing resources.  This analysis requires $\sim$1 CPU hour per 10 Monte Carlo iterations, so the $10^5$ iterations used in this analysis took multiple hours on almost a thousand cores of a high-performance computing cluster.

The spectral fraction analysis found results very similar to those from the mutual Hill radius measure, as the injection acceptance rates for each method was 1-2\%.  In addition, a very small amount of iterations of the system with an additional planet ($\lesssim 5\times10^{-4}$) were actually found to be unstable via the empirically-determined threshold values for the spectral fraction from the analysis by \citet[][]{Volk2020}.  This could likely be a feature of our crossing orbit limit; if a system survives through $10^6$ orbits but two planets have crossing orbits that are unstable from spectral fraction, the results count this as unstable via crossing orbits and not from the spectral fraction analysis.  If the empirical threshold (1\% of the total power spectrum is within 5\% of the peak value) is loosened slightly (i.e., 1\% of the total power spectrum is within 1\% of the peak value, as a majority of the power spectrum is multiple orders of magnitude below the peak), we see a modest increase in the fraction of injections found to be unstable via spectral fraction to $\sim 2\times10^{-3}$.  Therefore, it is likely that very few injections into the \hd{} system would allow for secular chaos to cause instabilities that would be noticeable in the spectral fraction analysis, even for the highest-multiplicity system architecture hypothesis H[a].

However, one significant quantitative difference is the effects of resonances.  In particular, in the results from Hypotheses H[a] and H[c] that include planet f, the N-body simulations and spectral fraction analysis find notable drops in the period relative likelihood at both the 2:1 and 3:2 resonances external to planet c and internal to planet f, as well as the overlapping 5:3 (with 2:1) and 11:6 (with itself) resonances.  This is consistent with the known pattern in the exoplanet population \citep[e.g.,][]{Petrovich2013,Ramos2017}.  However, there are notable examples of multi-planet resonant chains that are stable \citep[e.g.,][]{Gillon2017b,HardegreeUllman2021}.

\subsection{Lessons Learned on Dynamical Stability Analysis}

The mutual Hill radius dynamical stability criterion seems to be a good blend of accuracy and speed.  The results are similar to the N-body integration spectral fraction analysis within 0.5\% for a very large majority of cases where $\Delta \geq 10$.  Furthermore, even running $10^5$ Monte Carlo iterations only takes a few seconds on a standard 8-core desktop computer.  Thus, in general it would be the likely method of choice for many multi-planet systems.

However, if the system architecture near resonances is being tested, or if there are tightly packed planets where an injection is still deemed likely by the mutual Hill radius criterion, then an N-body integration method would likely be more useful.  SPOCK can be utilized to gain a quick understanding of the likely stability of the system before any additional planets are injected.  If the value SPOCK reports is near the stability threshold, then the spectral fraction analysis can be performed to determine the stability at a deeper level than the simple model; this requires more computational resources to be available.

For future studies, we will create a hybrid model that chooses which criterion to utilize for a given Monte Carlo iteration of the injected planet parameters.  If the distance between planets is within a certain value of the mutual Hill radius (larger than the current limit) or is near a low-order resonance, we would apply a method with N-body integrations.  For the majority of Monte Carlo iterations (those that lay outside of the mutual Hill radius and resonance constraints), we would use the simple stability criterion of the mutual Hill radius when adding in the additional planet.

\section{Discussion - Solar System Comparison} \label{sec:SS_disc}

\begin{figure*}[ht]
    \centering
    \includegraphics[width=1.05\columnwidth]{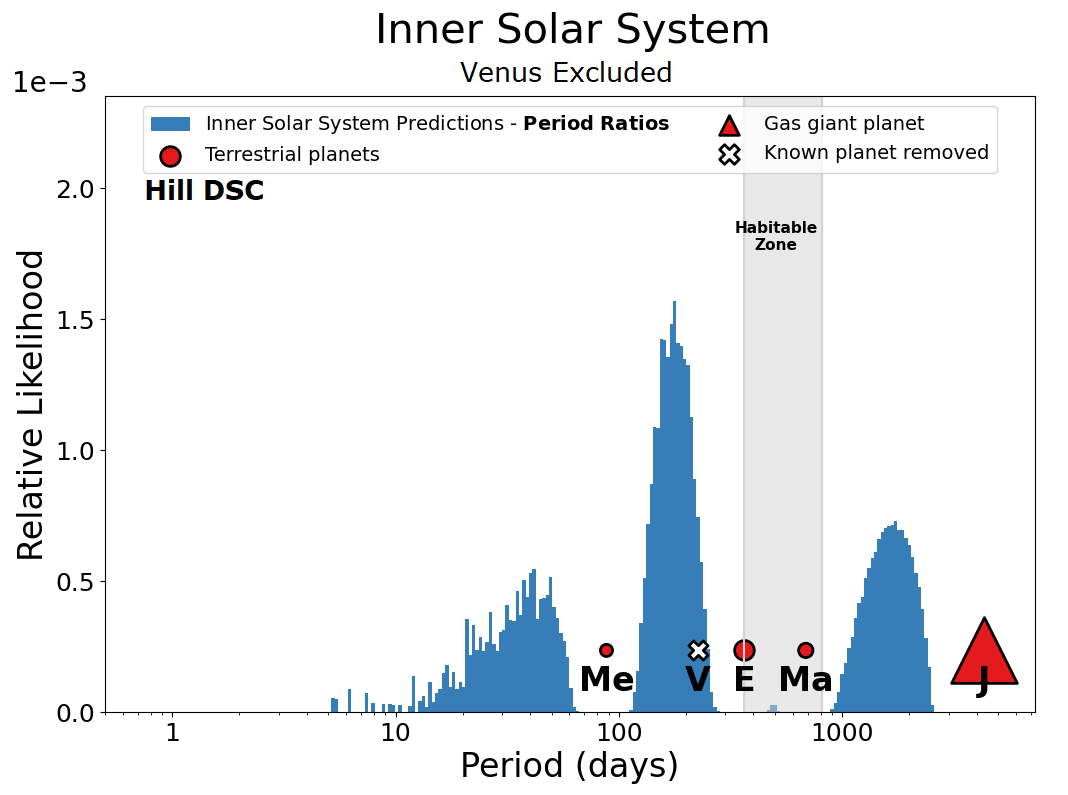}
    \includegraphics[width=1.05\columnwidth]{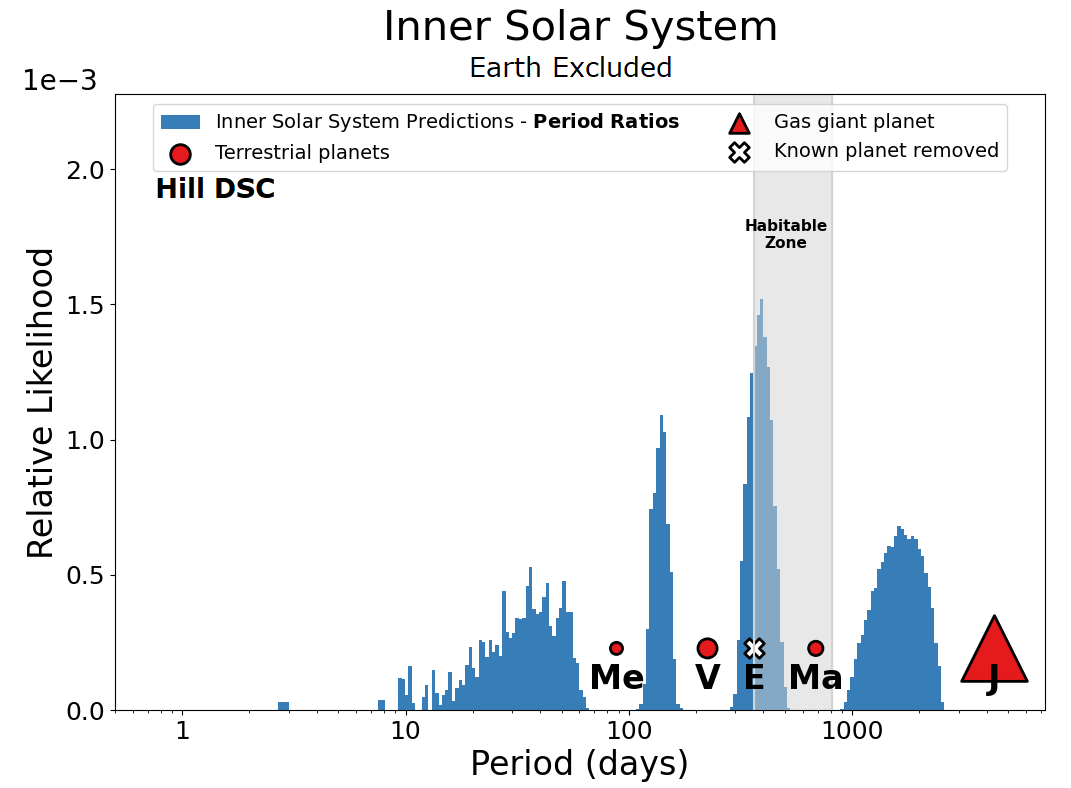}
    \includegraphics[width=1.05\columnwidth]{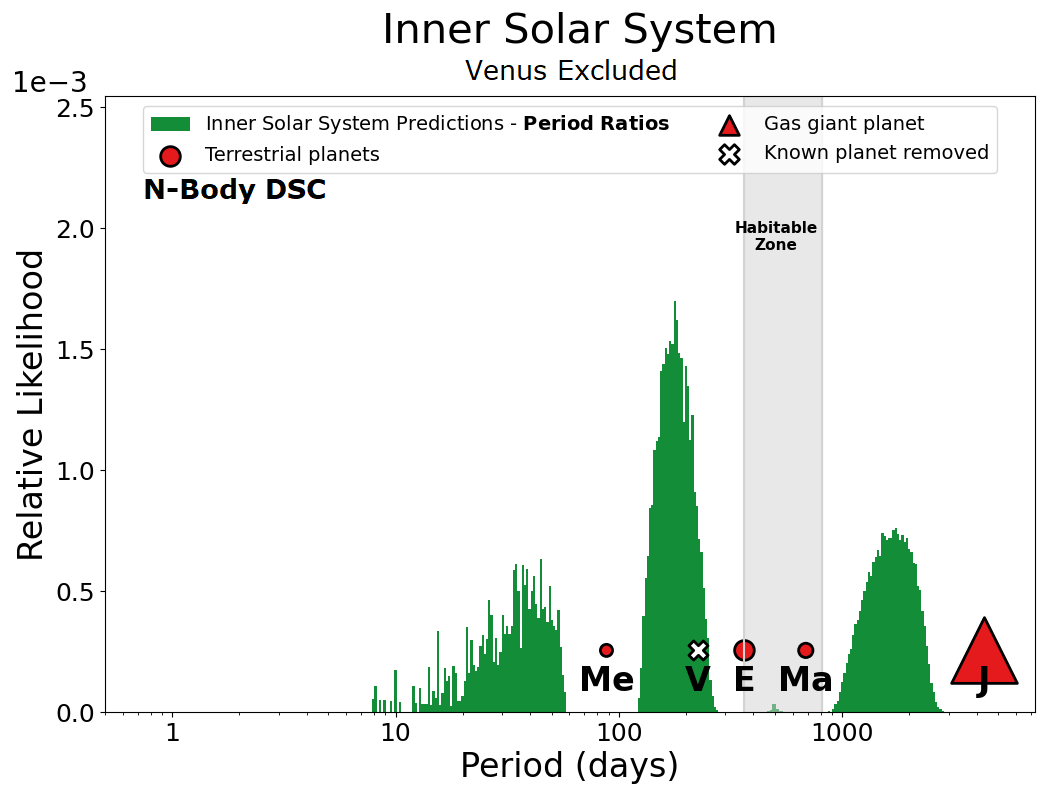}
    \includegraphics[width=1.05\columnwidth]{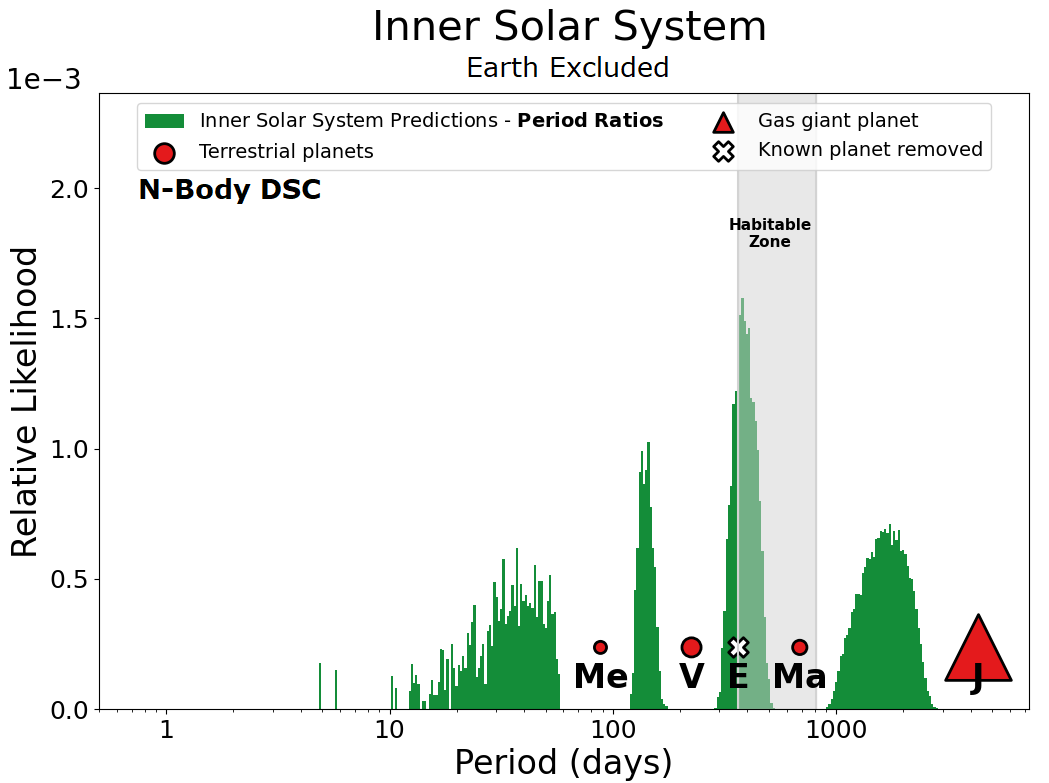}
    \caption{Top: The relative likelihoods in the inner Solar System (+ Jupiter) after removing a planet, utilizing the mutual Hill radius stability criterion. Bottom: The relative likelihoods after removing a planet utilizing the N-body spectral fraction analysis.  Left: Venus is excluded from the analysis, showing a large gap where a planet is likely to be located.  Right: Earth is excluded from the analysis, showing one prominent gap and one tight gap where a planet could be located.  All show a relatively large cumulative probability of a planet in the region of the Solar System occupied by the asteroid belt.}
    \label{fig:SS}
\end{figure*}

To compare the results from our different stability criteria for the best-known system, we tested our analysis on a few variants of the inner Solar System to check the performance.  We analyzed systems consisting of the terrestrial planets plus Jupiter, and tested excluding Venus or Earth from this group of planets.  As for our analysis of \hd{}, we used the period ratios model for the orbital periods, the volatile-rich power-law for the mass-radius relationship, and the SysSim multiplicity-dependent Lognormals for the inclination and eccentricity models.  Figure~\ref{fig:SS} shows the results from this comparison.

We find in this case again that there is very little difference in the mutual Hill radius measure vs the N-body integration methods, as both SPOCK and the spectral fraction analysis agree that the system would be long-term stable even with an additional planet injected.  All methods agree that an additional planet would likely be found interior to Mercury \citep[around 35-40 days, given that we are utilizing the equal period ratio model with a constraint on the innermost planet in a system likely being near a 12-day orbit;][]{Mulders2018}, and also between Mars and Jupiter at $\sim$1800 days.  Of course, this second peak in the relative likelihood roughly corresponds to the location of the asteroid belt in the Solar System, which provides another line of questioning that is beyond the scope of this study: could the location of a predicted planet actually harbor a debris disk or debris ring, and would it be visible in infrared imaging of the system?

When Venus is excluded from the analysis of the inner Solar System, there is an additional large peak in the relative likelihood at $\sim$190 days, slightly interior to the actual orbit of Venus at 224 days.  When Earth is excluded, there is a large peak at $\sim$390 days, slightly exterior to the actual orbit of Earth.  Interestingly, in the case where Earth is excluded, all dynamical stability methods predict a small but noticeable probability of injecting a planet between Mercury and Venus at $\sim$140 days.  However, given the eccentricity of Mercury, it is likely that any planet positioned at this spot in period space could cause Mercury and/or the other planet to become unstable.  \tnt{} also finds significant likelihood of another planet interior to the orbit of Mercury, although less likely than at the orbital period of the planet excluded from the analysis.  We do know that there are no planets interior to Mercury, but it is well known that in this aspect the Solar System is atypical \citep[e.g.,][]{Mulders2018}; observers from another planetary system who have determined similar exoplanet demographic data to us would likely be surprised to not find a planet within 88 days in the Solar System.

\begin{figure*}[ht]
    \centering
    \includegraphics[width=2.1\columnwidth]{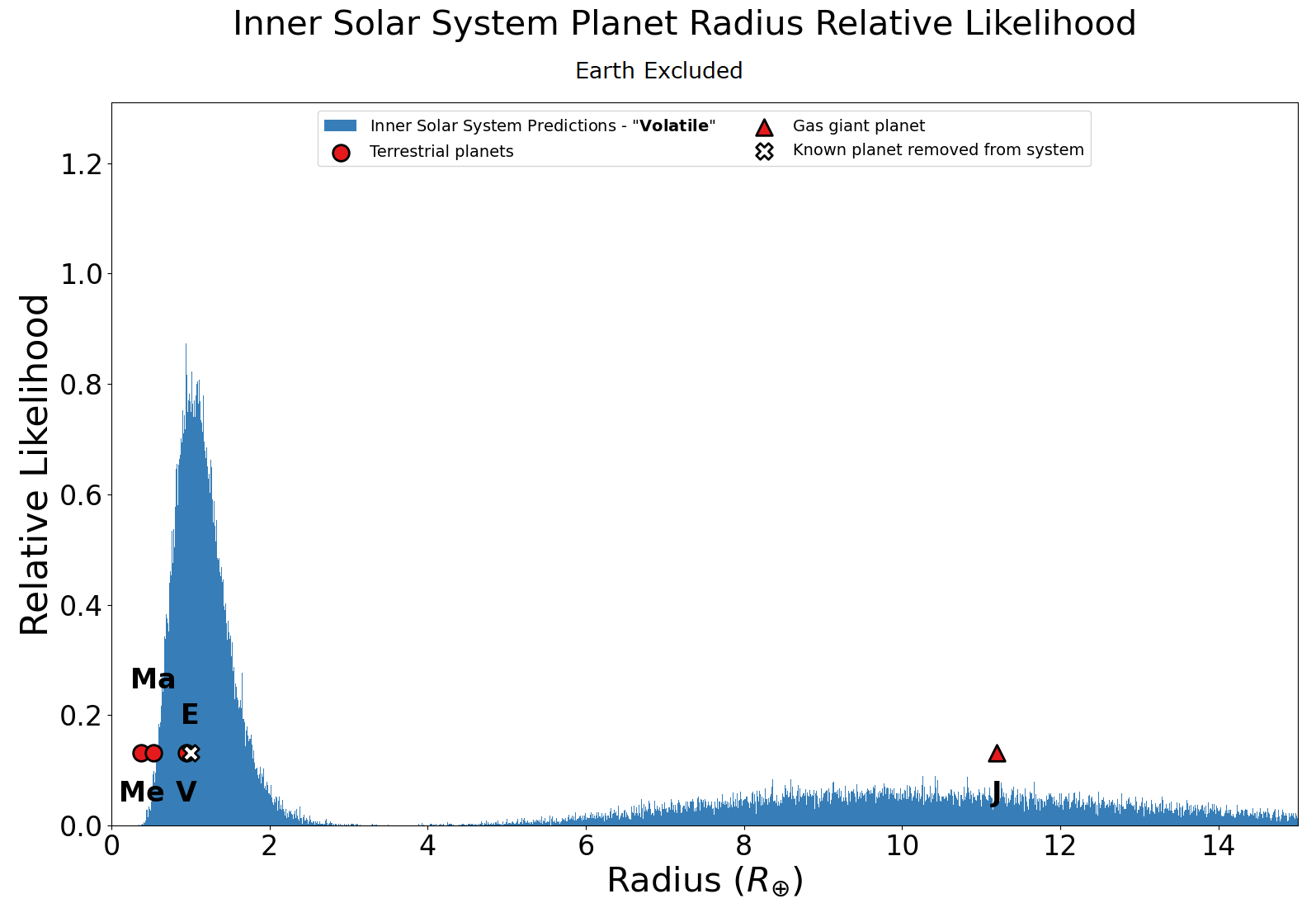}
    \caption{The relative likelihood for an additional planet's radius in the inner Solar System (+ Jupiter) after removing a planet, utilizing the mutual Hill radius stability criterion.  While some probability exists to see a larger planet, the peak is concentrated with the terrestrial planets.}
    \label{fig:SS_size}
\end{figure*}

The distribution for the predicted planet radius for additional planets in the Inner Solar System is shown in Figure~\ref{fig:SS_size}.  We find it is well matched to the terrestrial planets, since Jupiter is much larger than the rest of the planets and therefore would not be in the same ``cluster" of planet radius via the clustered radius model from SysSim \citep[][]{He2019}.  \tnt{} does find that there would be some probability of finding a planet with a comparatively large physical size, but the main peak in the planet radius distribution is centered on the size of the terrestrial planets.

In conclusion, our analysis is quite successful at predicting the presence of whichever planet is excluded from the inner Solar System, as well as the position of the asteroid belt corresponding to a potential additional planet.  While the Solar System is atypical as compared to the systems in the Kepler population on which our analysis is based \citep[e.g.,][]{Mulders2018}, the accurate predictions from \tnt{} on the Solar System offer another validation of our analysis.

\section{Summary} \label{sec:summary}

We present an integrated analysis of the \hd{} planetary system, the closest known 6-planet system, using an updated version of \tnt.  The key updates and findings of our study are as follows:

(1) We update and expand the \tnt{} software by allowing for unknown inclinations in the planet parameter distributions and including measurements of planet eccentricities.

(2) We present advanced dynamical stability criteria (including the inclinations and eccentricities) that utilize N-body integrations. One method uses machine learning to predict stability, and the other performs a ``spectral fraction" analysis of the angular momentum deficit (AMD) in the system.

(3) Utilizing both the simple mutual Hill radius dynamical stability criterion and the updated N-body integration spectral fraction analysis, in Hypotheses H[a] and H[c] (where planet f is considered to be real) we find evidence for predicted planet PxP--1 with orbital period $\sim$12 days (in between current planets c and f), which would dynamically pack the inner system out to planet g.

(4) Utilizing the same two dynamical stability criteria, in Hypotheses H[b] and H[d] (where planet f is considered to be uncertain) we find a large integrated probability spread out between the peak for PxP--1 in Hypotheses H[a] and H[c] and the expected position of planet f.  When PxP--1 is injected at 12 days and run \tnt{} again, we find that the new injections recover the period of planet f almost exactly.

(5) Using the SPOCK machine-learning predictions, we find that all hypotheses are considered stable without injections, and tends to find that including planet g makes the system more unstable in general, whereas including planet f increases the scatter in the likely stability.

(6) Using the SPOCK simple feature classifier, we find that the dynamical architecture of Hypothesis H[a] is likely fully packed and is likely unstable with any additional planets, while the other hypotheses allow room for additional planets.  Using the SPOCK deep regressor predictor, we find Hypotheses H[a] and H[b] are unstable with an additional planet, Hypothesis H[c] is very near the stability limit when injecting a planet, and Hypothesis H[d] is stable with injections.

(7) We find support for both ``controversial" planets, planet f with an orbital period of 22.7 days (very near the likely rotation period of its host star) and planet g with an orbital period of 94.2 days (only found in one study of the system through this year, much larger uncertainties).

(8) We also find support for a predicted planet PxP--2 external to the current inner system that would have an orbital period of $\sim$174 days, on the inner edge of the habitable zone, with a likely planet mass of $\sim4\:M_\oplus$.  Assuming dynamical packing and continuing outwards from there, another planet could be found at an orbital period of $\sim$321 days, in the outer parts of the habitable zone.  Long-term signals have been found in RV data at $\sim$192 days and $\sim$364 days, but are found to be more likely to be aliases from Earth's orbital period \citep[][]{Rosenthal2021}.

(9) We apply this integrated analysis to the inner Solar System up to and including Jupiter, and tested excluding one of the terrestrial planets from the input parameters.  Our predictions from the analysis are excellent matches to the known values from the Solar System planets, validating our results.

Our study demonstrates how additional information on dynamical stability can influence the understanding of system architectures and the possibility of finding additional ``hidden" planets in these systems.  The \hd{} system, being the closest system to our Sun with 6 planet candidates, was a prime case on which to test our \tnt{} algorithm's integrated analysis.  We find that dynamically complex systems (i.e., systems with high multiplicity, high eccentricity, etc.) would still allow the presence of another planet without losing dynamical stability.  Our study also provides support for the controversial planets/candidates in the \hd{} system and predictions for additional planets, one of which would lie in the inner part of the habitable zone.  This would make \hd{} an even more enticing target for direct imaging studies both from space and on the ground in the near future.

\acknowledgments

The results reported herein benefited from collaborations and/or information exchange within the program “Alien Earths” (supported by the National Aeronautics and Space Administration under Agreement No. 80NSSC21K0593) for NASA’s Nexus for Exoplanet System Science (NExSS) research coordination network sponsored by NASA’s Science Mission Directorate.  RM additionally acknowledges funding from NASA grant 80NSSC18K0397.  This research has made use of the NASA Exoplanet Archive and the Exoplanet Follow-up Observation Program website, which is operated by the California Institute of Technology, under contract with the National Aeronautics and Space Administration under the Exoplanet Exploration Program.  We acknowledge use of the software packages NumPy \citep[][]{Harris2020}, SciPy \citep[][]{Virtanen2020}, Matplotlib \citep[][]{Hunter2007}, and REBOUND \citep[][]{Rein2012, Rein2019}.  This paper includes data collected by the Kepler mission.  Funding for the Kepler mission is provided by the NASA Science Mission Directorate.  An allocation of computer time from the UA Research Computing High Performance Computing (HPC) at the University of Arizona is gratefully acknowledged.  The citations in this paper have made use of NASA’s Astrophysics Data System Bibliographic Services.

\bibliography{main}

\appendix{}

\section{\tnt{} Injection Figures} \label{app:figures}

\begin{figure*}[ht]
    \centering
    \includegraphics[width=0.49\columnwidth]{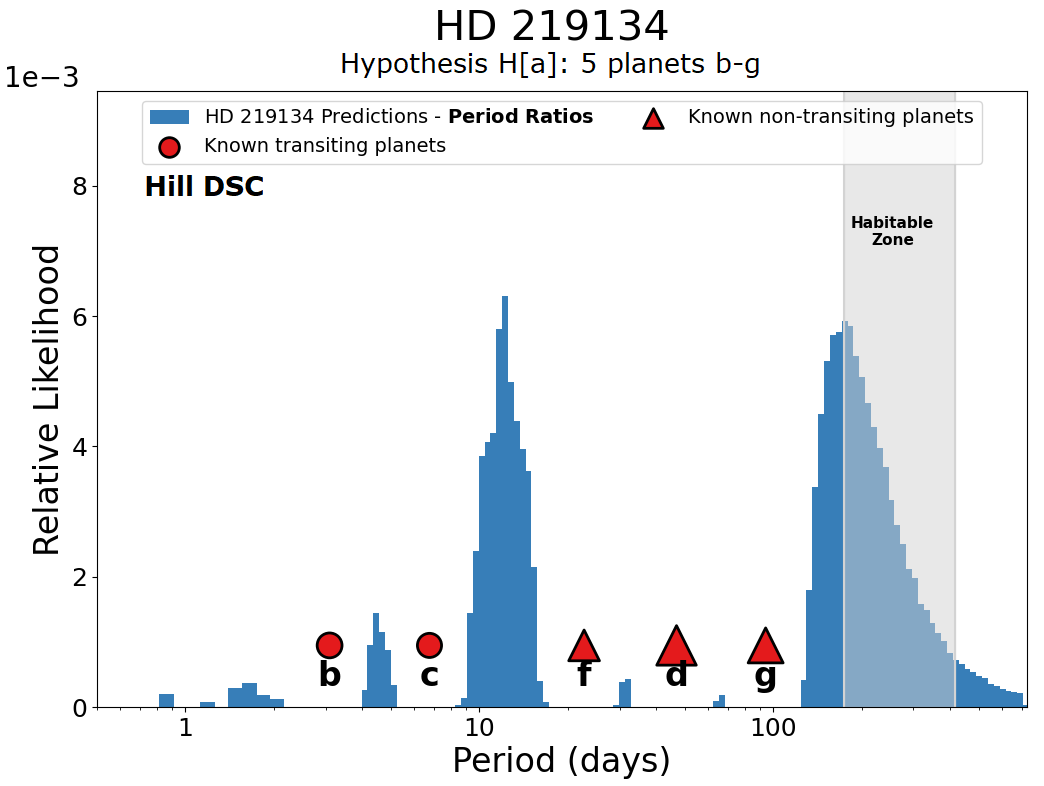}
    \includegraphics[width=0.49\columnwidth]{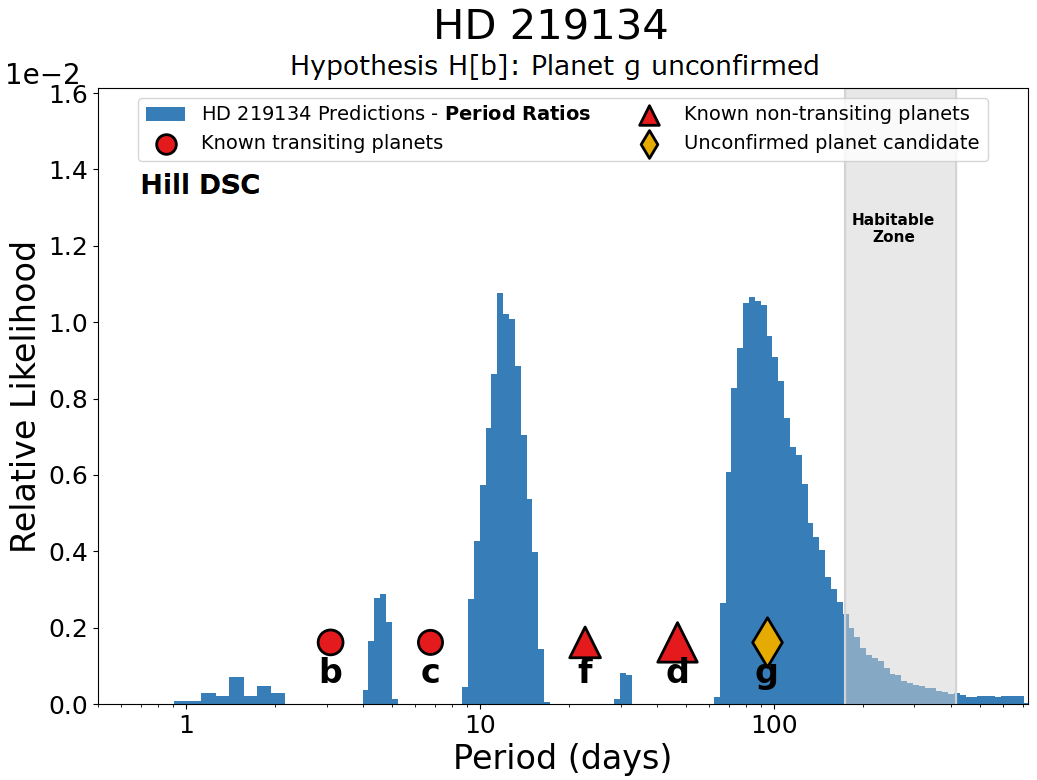}\\
    \vspace{10pt}
    \includegraphics[width=0.49\columnwidth]{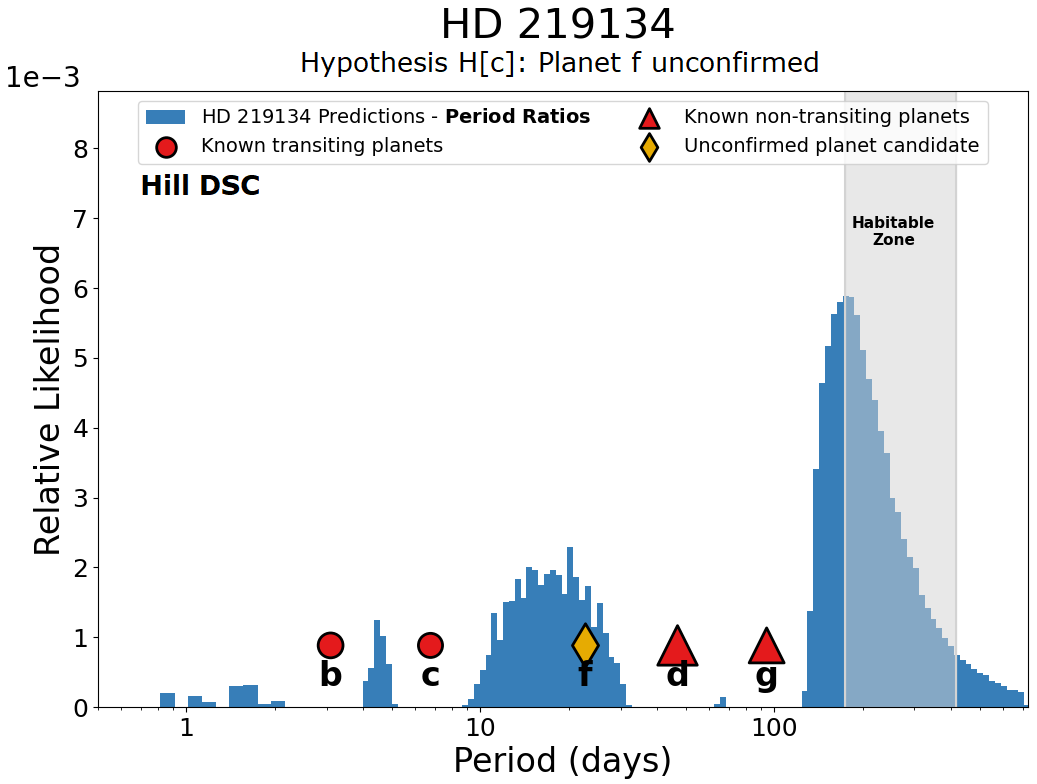}
    \includegraphics[width=0.49\columnwidth]{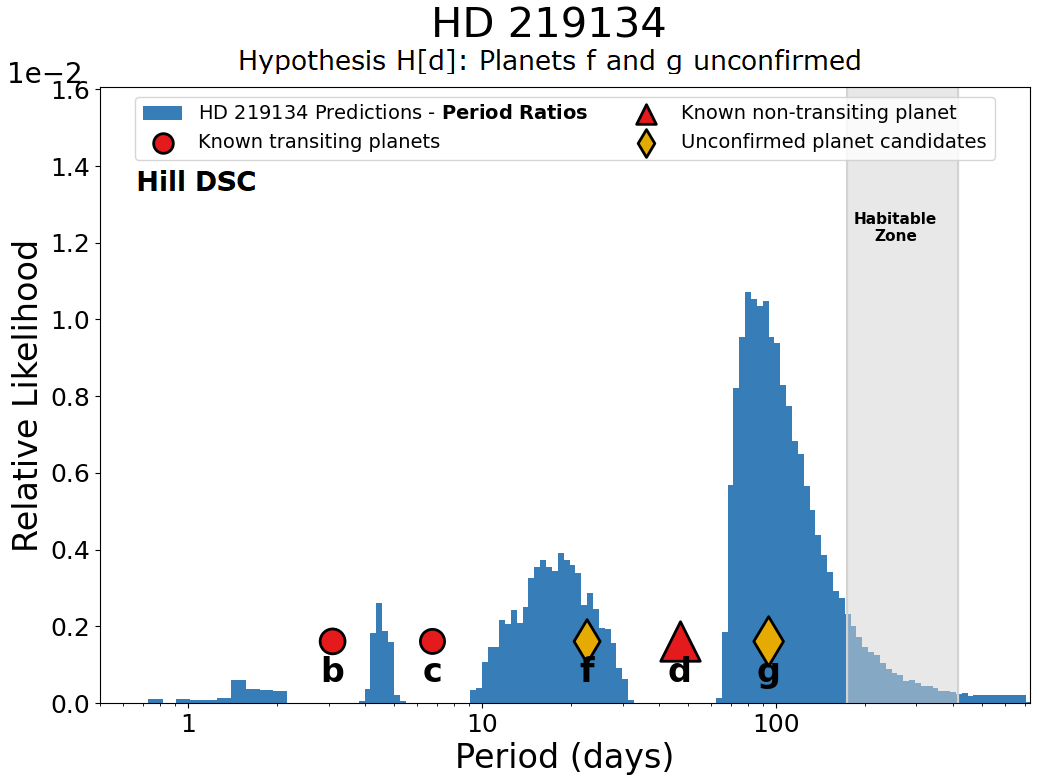}
    \caption{\tnt{} analysis utilizing the simple dynamical stability criterion requiring neighboring planets' separation in semimajor axis to exceed 8 mutual Hill radii.  Top left: Analysis assuming Hypothesis a, the full 5-planet system. Top right: Analysis assuming Hypothesis b, with no planet f (orbital period near stellar rotation period). Bottom left: Analysis assuming Hypothesis c, with no planet g (not found in all studies). Bottom right: Analysis assuming Hypothesis d, with no planet f nor planet g.  Excluding planet f causes the predictions to spread out in period space, indicative of two planets possibly missing.  Excluding planet g moves the exterior injection space to center almost directly on the period of planet g.}
    \label{fig:simplefg}
\end{figure*}

\begin{figure*}[ht]
    \centering
    \includegraphics[width=0.49\columnwidth]{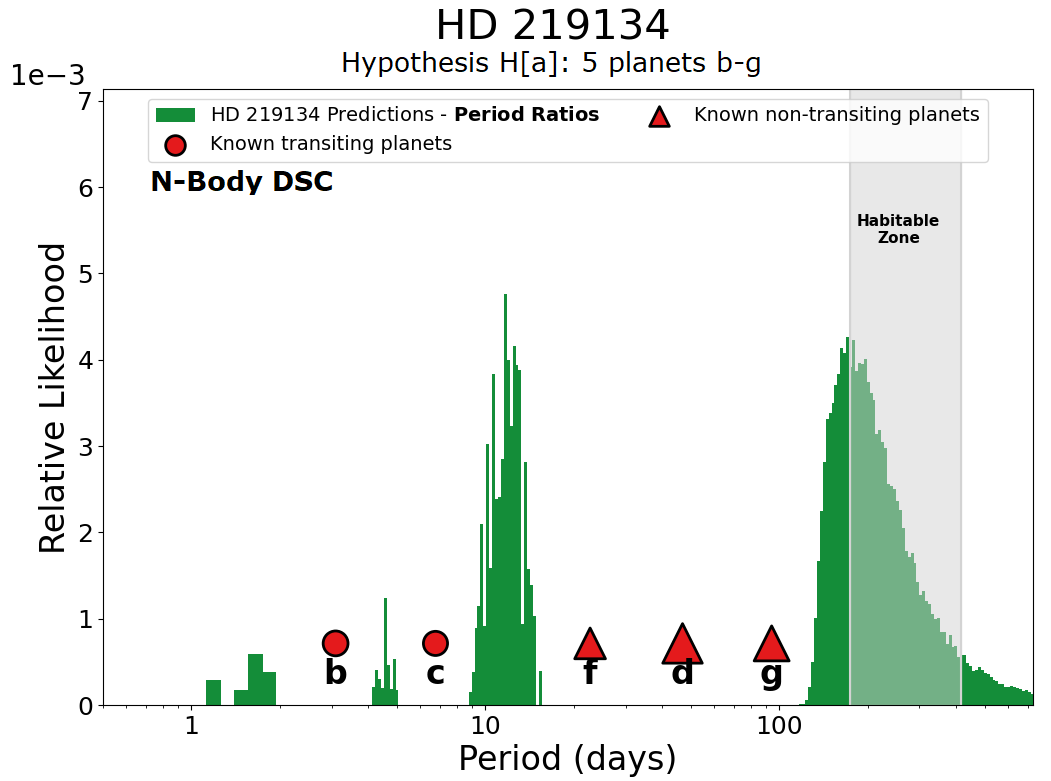}
    \includegraphics[width=0.49\columnwidth]{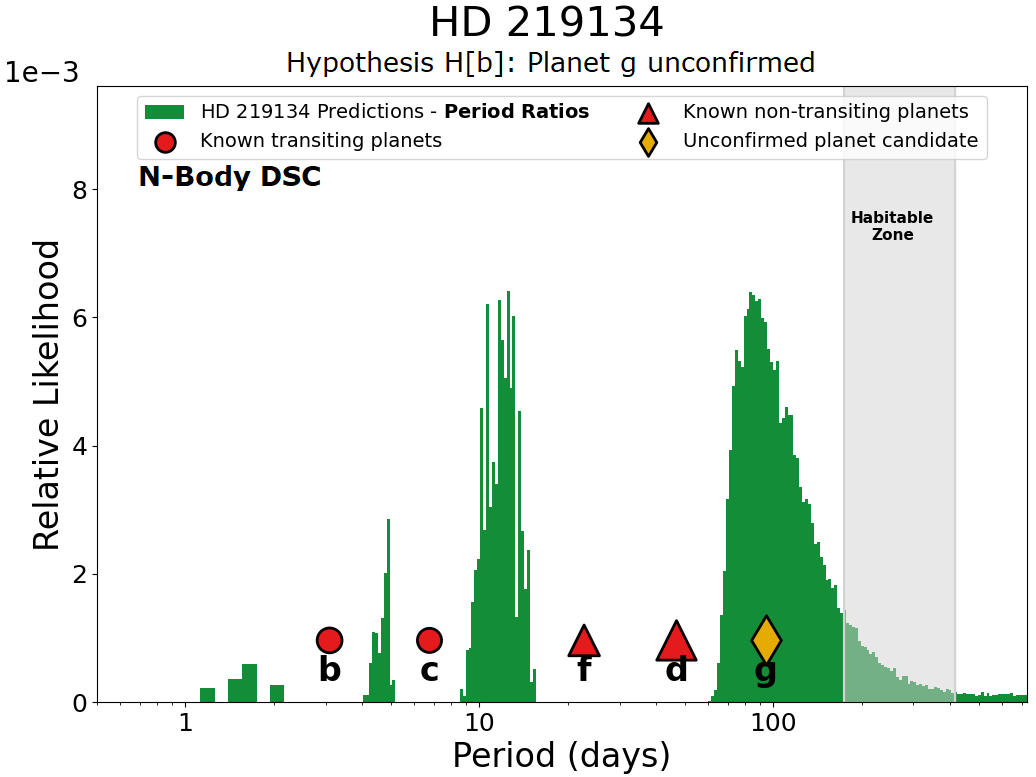}\\
    \vspace{10pt}
    \includegraphics[width=0.49\columnwidth]{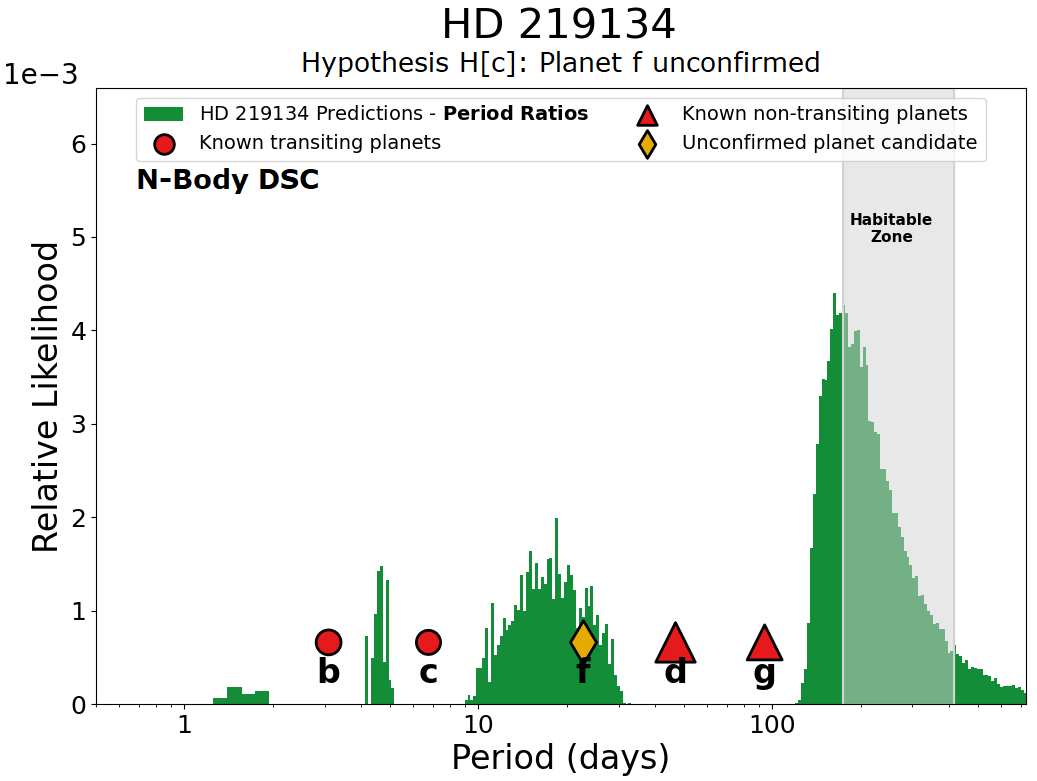}
    \includegraphics[width=0.49\columnwidth]{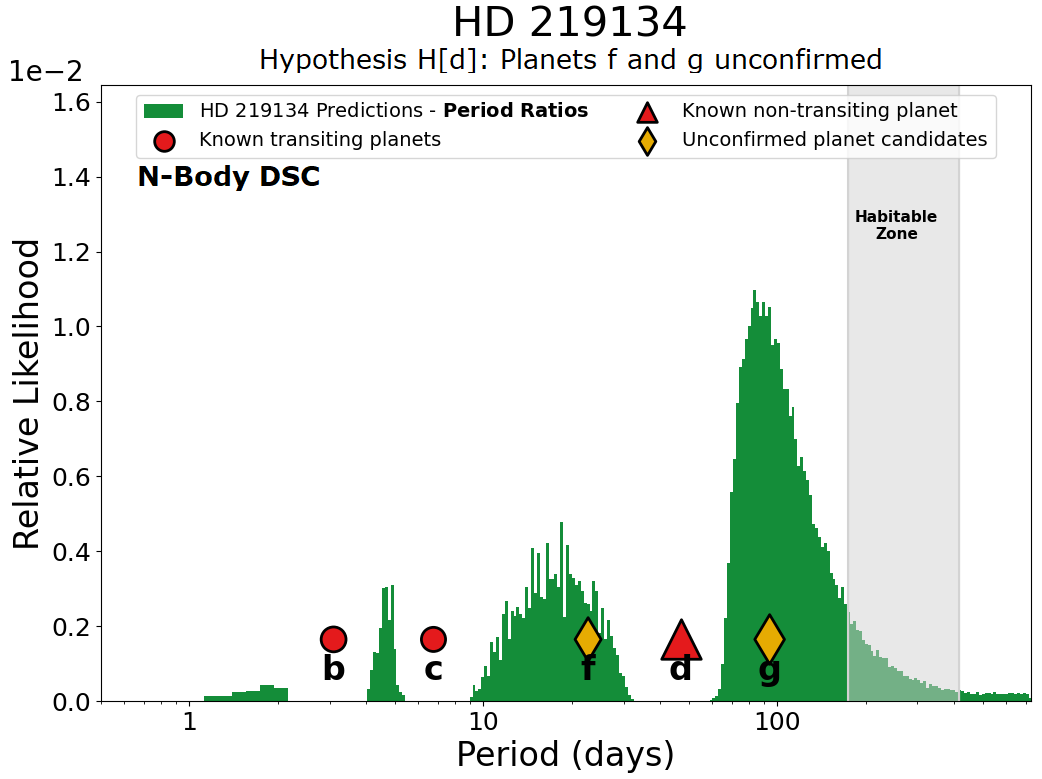}
    \caption{\tnt{} analysis utilizing the N-body simulations and testing orbit crossing and the AMD spectral fraction.  Top left: Analysis assuming full 5-planet system. Top right: Analysis assuming no planet f (orbital period near stellar rotation period). Bottom left: Analysis assuming no planet g (not found in all studies). Bottom right: Analysis assuming no planet f nor planet g.  The results here are similar to the simple dynamical stability criterion, only with tighter restraints near the $\Delta = 8$ limit and at orbital resonances.}
    \label{fig:dsc_test1}
\end{figure*}

\end{document}